\documentclass[twocolumn, trackchanges]{aastex701}

\usepackage{CJKutf8}
\usepackage{microtype}
\usepackage[T1]{fontenc}
\usepackage{amsmath}
\usepackage{multirow}
\usepackage{bm}
\usepackage{dsfont}
\usepackage{mathrsfs}
\usepackage[dvipsnames]{xcolor}
\usepackage{fontawesome}
\usepackage{graphicx}
\usepackage{booktabs}
\usepackage{placeins}
\usepackage[figure, figure*]{hypcap}
\usepackage{todonotes}
  \setuptodonotes{
    backgroundcolor=lightgray!50,
    linecolor=lightgray,
    bordercolor=none,
    tickmarkheight=0.2em,
  }

\newcommand{\gkai}[1]{\begin{CJK*}{UTF8}{gkai}\raisebox{.1em}{(}#1\raisebox{.1em}{)}\end{CJK*}}

\DeclareMathOperator*{\E}{\mathds{E}}
\DeclareMathOperator{\KL}{KL}
\DeclareMathOperator{\tprod}{{\textstyle\prod}}

\newcommand{\R}{\mathds{R}}
\newcommand{\Z}{\mathds{Z}}

\renewcommand{\d}{\mathrm{d}}
\newcommand{\p}{\partial}
\newcommand{\mVert}{\mathrel{\Vert}}
\bmdefine{\vzero}{0}
\bmdefine{\vI}{I}
\bmdefine{\vx}{x}
\bmdefine{\vz}{z}
\bmdefine{\vtheta}{\theta}
\newcommand{\Loss}{\mathscr{L}}
\newcommand{\batch}{\hat{B}}
\newcommand{\dmo}{\mathrm{dmo}}
\newcommand{\hyd}{\mathrm{hyd}}

\newcommand{\Mpc}{\mathrm{Mpc}}

\newcommand{\Y}[1]{\textcolor{Bittersweet}{\textbf{#1}}}

\begin{document}

\title{One latent to fit them all: a unified representation of baryonic
feedback on matter distribution}
\shorttitle{One latent to fit them all}
\submitjournal{ApJL}

\author[0009-0000-5381-7039]{Shurui Lin \gkai{林书睿}}
\affiliation{Department of Astronomy, University of Illinois at
Urbana-Champaign, 1002 West Green Street, Urbana, IL 61801, USA}
\email[show]{shuruil3@illinois.edu}

\author[0000-0002-0701-1410]{Yin Li \gkai{李寅}}
\affiliation{Department of Strategic and Advanced Interdisciplinary
Research, Peng Cheng Laboratory, Shenzhen, Guangdong 518000,
China}
\email[show]{eelregit@gmail.com}
\correspondingauthor{Shurui Lin \& Yin Li}

\author[0000-0002-3185-1540]{Shy Genel}
\affiliation{Center for Computational Astrophysics, Flatiron Institute,
162 Fifth Avenue, New York, NY 10010, USA}
\affiliation{Columbia Astrophysics Laboratory, Columbia University, 550
West 120th Street, New York, NY, 10027, USA}
\email{shygenelastro@gmail.com}

\author[0000-0002-4816-0455]{Francisco Villaescusa-Navarro}
\affiliation{Center for Computational Astrophysics, Flatiron Institute,
162 Fifth Avenue, New York, NY 10010, USA}
\affiliation{Department of Astrophysical Sciences, Princeton University, 4 Ivy Lane, Princeton, NJ 08544 USA}
\email{villaescusa.francisco@gmail.com}

\author[0000-0003-2734-0294]{Biwei Dai \gkai{戴必玮}}
\affiliation{School of Natural Sciences, Institute for Advanced Study, 1
Einstein Drive, Princeton, NJ 08540, USA}
\email{biwei@ias.edu}

\author[0000-0003-1297-6142]{Wentao Luo \gkai{罗文涛}}
\affiliation{School of Aerospace Information, Hefei Institute of
Technology, 6 Zhizhong Road, Hefei, Anhui 238706, China}
\email{wentao.luo82@gmail.com}

\author[0000-0002-1512-5653]{Yang Wang \gkai{汪洋}}
\affiliation{Department of Strategic and Advanced Interdisciplinary
Research, Peng Cheng Laboratory, Shenzhen, Guangdong 518000,
China}
\email{wangy18@pcl.ac.cn}

\begin{abstract}

Accurate and parsimonious quantification of baryonic feedback on
matter distribution is of crucial importance for understanding both
cosmology and galaxy formation from observational data.
This is, however, challenging given the large discrepancy among
different models of galaxy formation simulations, and their distinct
subgrid physics parameterizations.
Using 5,072 simulations from 4 different models covering broad ranges in their
parameter spaces, we find a unified 2D latent representation.
Compared to the simulations and other phenomenological models, our
representation is independent of both time and cosmology, much
lower-dimensional, and disentangled in its impacts on the matter power
spectra.
The common latent space facilitates the comparison of parameter spaces of different models and is readily interpretable by correlation with each.
The two latent dimensions provide a complementary representation of baryonic effects,
linking black hole and supernova feedback to distinct and interpretable impacts on
both the matter power spectrum, and field, level.
Our approach enables developing robust and economical analytic models for optimal gain of physical information from data, and is generalizable to other fields with significant modeling uncertainty.

\end{abstract}


\section{Introduction}
\label{sec:intro}

Observational cosmology is entering a new era, marked by the rapid
advancement of Stage-IV experiments, including Euclid
\citep{EuclidQ1_Overview}, The Dark Energy Spectroscopic Instrument (DESI) \citep{DESIDR2_cosmology}, Legacy Survey of Space and Time (LSST) performed by Vera C. Rubin Observatory
\citep{Ivezić2019LSST}, and Nancy Grace Roman Space Telescope (Roman) \citep{2024Roman}.
While most cosmological observables focus on the luminous baryonic
matter in our Universe, weak gravitational lensing additionally has the unique
strength in directly probing the total matter distribution
\citep{Kilbinger2015WL}, that of both baryonic and dark matter.
Optimal analyses of weak-lensing data require
accurate modeling of baryonic physics on small scales, where galaxies form
and feed back to the environment \citep{Lu2021small_scale}.
However, exploiting small-scale weak-lensing data requires accurate baryonic physics models up to $k_\mathrm{max}\sim 0.3\ h/\mathrm{Mpc}$ \citep{LSST_WL_Requirements},
and quantitative agreement remains a significant obstacle, limiting our
ability to make precise inferences and fully exploit the potential of
forthcoming surveys \citep{Vogelsberger2019sim, Copeland2018de, Huang2019bar}.
Therefore, an accurate model of
baryonic effects at small scales is crucial to achieving precise cosmological measurements.

To study the non-linear structure formation and multi-scale galaxy
formation, numerical simulations, especially hydrodynamical simulations,
have become the essential tool.
Hydrodynamical simulations model gas dynamics, star formation, and feedback mechanisms \citep{Di_Matteo2005feedback, Salcido2023feedback},
with notable examples including IllustrisTNG, Astrid, Swift-EAGLE, and SIMBA \citep{nelson2021TNG, Dav2019SIMBA, Bird2022Astrid, Crain2015EAGLE},
offering a broad spectrum of feedback models and parameterizations.
Building upon these developments, the Cosmology and Astrophysics with
Machine Learning Simulations (CAMELS) project has recently generated an
extensive set of cosmological simulations that systematically vary both
cosmological and baryonic physics parameters across different feedback
models \citep{Paco2021CAMELS, Ni2023SB28}.
This rich dataset enables a comprehensive investigation of baryonic
feedback effects using advanced statistical and machine learning
techniques.

Hydrodynamical simulations are computationally expensive, typically
costing more than one order of magnitude than the $N$-body simulations
that evolve total matter under only gravity.
This high cost severely limits the ability to fully sample cosmological
and astrophysical parameter spaces,
creating a bottleneck for accurate modeling of baryonic feedback
in large-scale structure analyses.
Fortunately, baryonic feedback primarily
alters the amplitude rather than the phase of Fourier modes
in the matter density field,
leaving the mode phases largely unchanged, even on small scales
($k \lesssim 10\ h\mathrm{Mpc}^{-1}$) across different hydrodynamical simulations \citep{SharmaEtAl2024}.
Instead of evolving the hydrodynamical density field fully,
one can focus on modeling how feedback modifies
the amplitude of density fluctuations relative to a dark-matter-only baseline.
A convenient way to encode this effect is through the squared transfer function ($T^2$)
defined by the ratio of the hydrodynamic power spectra ($P_\hyd$)
and dark-matter-only power spectra ($P_\dmo$):
\begin{equation}
T^2(k, a) = \frac{P_\hyd(k, a)}{P_\dmo(k, a)},
\end{equation}
which enables efficient modeling of baryonic feedback
at \emph{both the field level and the power-spectrum level} of total matter distribution
simply from the $N$-body runs,
avoiding the cost of large hydrodynamic suites while retaining feedback effects.

With the growing availability of simulation datasets and advances in machine learning,
several recent studies have modeled the transfer function $T^2$ via
Emulation-based methods
\citep{SharmaEtAl2024, zhou2025emulator, Schaller2025Flamingo},
analytic formulations,
\citep{Schaller2025anl_bar, kammerer2025sr}
and physically motivated parameterizations
 \citep{Medlock2025FRB,vanDaalen2019}.
Alternatively to hydrodynamical simulation, baryonification or baryon correction models apply parametrized prescriptions to displace matter distribution in gravity-only simulations to account for the influence of baryons \citep{Aric2020Feedback, Aric2021BACCO, Aricò2021feedback}.
Despite their successes, most approaches are limited to
a single simulation model or a narrow parameter space,
which can hinder generalization.
Moreover, many introduce multiple, often redshift-dependent, parameters,
leading to degeneracies that reduce information content and obscure physical interpretability.
Collectively, these examples underscore the difficulty of integrating data from different simulation suites with a wide range of parameter choices
and the challenge of developing a disentangled representation for baryonic feedback modeling.
We also provide a summarized comparison in \autoref{sec:compare_methods}.

With the development of machine learning,
variational autoencoders (VAEs) \citep{kingma2013VAE, Rybkin2021simple}
offer a way to learn latent representations adaptable to multiple datasets \citep{Lucie_Smith2022VAEhalo, Piras2024w0wa}.
Furthermore, to improve the interpretability of the learned representations,
\cite{burgess2018understanding} introduced the $\beta$-VAE,
which encourages disentanglement in the latent space
by increasing the weight on the KL divergence term.
However, this often comes at the cost of reduced reconstruction accuracy
and limited information capacity in the latent representation.
\cite{chen2019isolating} proposed the $\beta$-TCVAE,
which incorporates additional hyperparameters to upweight the total correlation term in the loss function
(see \autoref{sec:VAEs}) to explicitly disentangle the latent dimensions
while preserving both reconstruction quality and information in the latent space.

In this paper, we propose a $\beta$-TCVAE–based representation-learning approach to reconstruct the transfer function $T^2$ and learn a latent representation of baryonic feedback across four CAMELS simulation suites over $0.356<k<10,h/\mathrm{Mpc}$ and $0\le z\le2$.
Based on these, we achieve the following major contributions:
\begin{itemize}
  \item \textbf{Nonlinearity}:
  The neural-network and variational capacity of our VAE model captures the non-linear dependence of baryonic feedbacks on galaxy formation physics, across thousands of simulations and multiple subgrid physics models without amplifying noise.

  \item \textbf{Independence}:
  The latent space is trained to be independent of both redshift and cosmology, preventing duplicating the cosmological dependence and avoiding multiplying the latent dimensionality with number of redshift bins.

  \item \textbf{Parsimony}:
  The latent space is only 2D, minimizing information loss when marginalized over in cosmological inference.

  \item \textbf{Interpretability}:
  The two latent parameters exhibit distinct scale and time dependences, and associate with different feedback mechanisms by correlation with galaxy formation parameters.
\end{itemize}

This paper is organized as follows:
In \autoref{sec:data}, we introduce the datasets we used in this study.
In \autoref{sec:VAE_main}, we introduce the design of the VAE model and loss.
We show the training strategy in
\autoref{sec:training}.
In \autoref{sec:result}, we show the results of the reconstruction of
spectrum ratio and the latent representation of baryonic feedback
effects.
In \autoref{sec:Discussion}, we discuss the results and the potential
application of the latent representation.
The appendices include more details, including, notably, a comparison with the principal component analysis (PCA) approach in \autoref{app:PCA}.

\section{Methods}
\label{methods}

\subsection{Data}
\label{sec:data}

The \textit{Cosmology and Astrophysics with MachinE Learning Simulations} (CAMELS) project
comprises over 15{,}000 cosmological simulations,
including 8{,}925 hydrodynamic and 6{,}136 N-body runs.
In this paper, we utilize the first-generation simulations of CAMELS,
each performed in a periodic box of $25~h^{-1}\,{\rm Mpc}$ per side,
tracking the evolution of $256^3$ dark matter particles and $256^3$ initial fluid elements.
This setup defines the baseline resolution for this simulation suite.
Our analysis focuses on four specific hydrodynamic suites\footnote{\url{https://camels.readthedocs.io/en/latest/parameters.html}}
, IllustrisTNG SB28, Astrid SB7, SIMBA LH6, and Swift-EAGLE LH6,
each characterized by variations in cosmological and astrophysical parameters,
as well as distinct random seeds for initial conditions.
We summarize the basic information of the four suites in
\autoref{table:CAMELS} in \autoref{sec:camels}.

To assess the robustness of our results, we additionally incorporate the CAMELS \textit{Cosmic Variance} (CV) set for benchmarking.
This set comprises 27 simulations per suite, each sharing the same fiducial cosmological and astrophysical parameters but differing in the initial condition random seed.
These simulations are commonly used to isolate the effects of cosmic variance.
We use the CV counterparts of our main four suites: IllustrisTNG CV, Astrid CV, SIMBA CV, and Swift-EAGLE CV.
For a detailed description of the above simulations, we refer the reader to \cite{Paco2021CAMELS, Ni2023SB28}.

For all suites, each hydrodynamic simulation is
paired with a gravity-only simulation (which is also called a
dark-matter-only simulation) with the same initial conditions.
We compute the transfer function $T^2$ from
the ratio of hydrodynamic to dark-matter-only total power spectra.
Data are split into training, validation, and test sets in an $1/2 \ {:} \ 7/16 \ {:} \ 1/16$ ratio
to ensure robust values for the KL terms in validation loss (see \autoref{sec:minibatch}).
From each simulation, we use three snapshots ($a \approx 1/3, 1/2, 1$),
rebin the power spectra into 18 $k$-bins over
$k \in [0.356, 28.90] h\mathrm{Mpc}^{-1}$
and train the model with the $3\times18$ dimensions spectrum ratio.
Based on the requirement of LSST \citep{LSST_WL_Requirements}
and the noise level in the large-$k$ end,
we limit our inference to $0.356<k<10\ h/\Mpc$.

\subsection{Variational autoencoders}
\label{sec:VAE_main}

In this section, we breifly review Variational Autoencoders (VAEs)
and outline our model design.

Variational Autoencoders (VAEs) are latent-variable models consisting of an encoder,
a probabilistic bottleneck, and a decoder \citep{kingma2013autoencoding}.
The encoder approximates the posterior over latent variables,
and the decoder reconstructs data from sampled latents.
Training balances reconstruction accuracy with a regularization term enforcing a prior.
To improve interpretability, $\beta$-VAE introduces a tunable weight on the regularization term
for more disentangled representations \citep{burgess2018understanding}.
$\beta$-TCVAE further decomposes the regularization into three
components to encourage decorrelated probability distribution among the
latents, as this has been found to promote their semantic
disentanglement \citep{chen2019isolating}.

We develop a conditional $\beta$-TCVAE model with the Convolutional Neural Network (CNN)
to reconstruct the power spectra ratio $T^2(k, a)$,
aiming to isolate the baryonic effect within the latent space,
while minimizing the influence of cosmological parameters.
Simultaneously, we aim for a disentangled latent space. This entails not just ensuring the latent variables are statistically independent, but also that each latent variable captures a unique feedback mechanism pattern.
For this, the model is conditioned on five cosmological parameters
\footnote{The total matter density $\Omega_\mathrm{m}$,
the amplitude of matter fluctuations $\sigma_8$,
the baryon density $\Omega_\mathrm{b}$,
the spectral index of the primordial power spectrum $n_\mathrm{s}$,
and the dimensionless Hubble parameter $h$}.
The top panel of \autoref{fig:AIO_plot} shows the scheme of the model.

To obtain a redshift-independent latent,
we consider three $T^2$ of snapshots from
the same simulation ($a \approx \tfrac{1}{3}, \tfrac{1}{2}, 1$)
as one single input.

The encoder takes $T^2$ with the five cosmological parameters
and outputs latent distribution:
\begin{equation}
\vz \sim q\bigl( \vz \mid T^2(k, a), {\Omega_\mathrm{m},
                \Omega_\mathrm{b}, \sigma_8, n_\mathrm{s}, h} \bigr),
\label{eq:encoder}
\end{equation}
with scale factor $a=\tfrac{1}{3}, \tfrac{1}{2}, 1$.

The decoder then reconstructs $T'^2$ from the latent and cosmological parameters:
\begin{equation}
{T'}^2(k, a) \sim p\bigl( {T'}^2(k, a) \mid \vz, {\Omega_\mathrm{m},
                  \Omega_\mathrm{b}, \sigma_8, n_\mathrm{s}, h} \bigr).
\label{eq:decoder}
\end{equation}
Our evidence lower bound (ELBO) loss function comprises four terms,
designed to ensure both high-quality reconstructions
and informative and disentangled latent representation:
\begin{itemize}
    \item \textbf{Reconstruction loss}: Measures the quality of the model's reconstruction of the input data.
    \item \textbf{Mutual information loss (MI-loss)}:
    The first KL term, corresponding to the mutual information between the latent variables and the samples.
    \item \textbf{Total correlation loss (TC-loss)}:
    The second KL term, promoting a disentangled latent representation.
    If this term vanishes, the latent distribution factorizes over its 1D marginals.
    \item \textbf{Dimension-wise KL loss (dw-KL-loss)}:
    Constrains each latent dimension to follow the prior distribution, typically a standard normal.
\end{itemize}
Readers can find more details in \autoref{sec:VAE}.

\subsection{Training strategy}
\label{sec:training}

We use the Python package \texttt{Optuna} to optimize the
hyperparameters of the model \citep{Optuna}.
Details about hyperparameter tuning and model selection are described in \autoref{app:hpara}.

We trained five different models.
Four of them are single-dataset models, trained on IllustrisTNG SB28, Astrid SB7, Swift-EAGLE LH6
and SIMBA LH6, as mentioned in \autoref{sec:data}.
The fifth model is a model trained on a general dataset that
includes IllustrisTNG, Astrid, and Swift-EAGLE, denoted as
``TEA'' in the following part of this paper.
SIMBA was not included in the general dataset as its feedback model
is too strong compared with the other three datasets
(readers can find more about this in \autoref{app:EAST}).
We do our test on all five different kinds of models to see if the
latent space is consistent across different datasets and models.
The TEA model is trained on the three datasets with the different
cosmology and feedback models, so it should be able to learn a more
general representation of the baryonic feedback effects.

\section{Results}
\label{sec:result}

\subsection{Reconstruction}
\label{sec:rec}

\begin{figure*}[tb]
  \centering
  \includegraphics[width=0.8\textwidth]{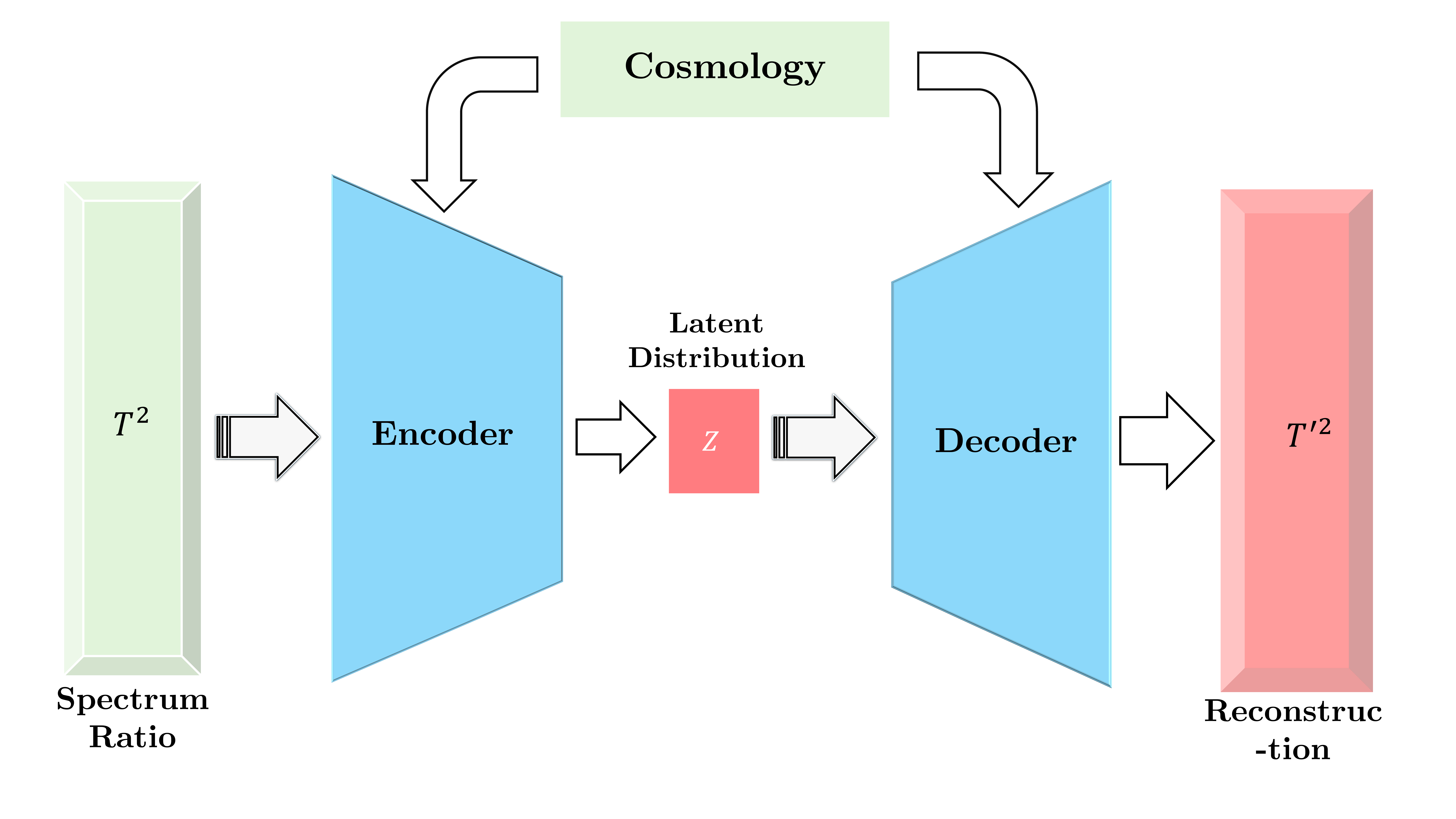}
  \includegraphics[height=0.4\textwidth]{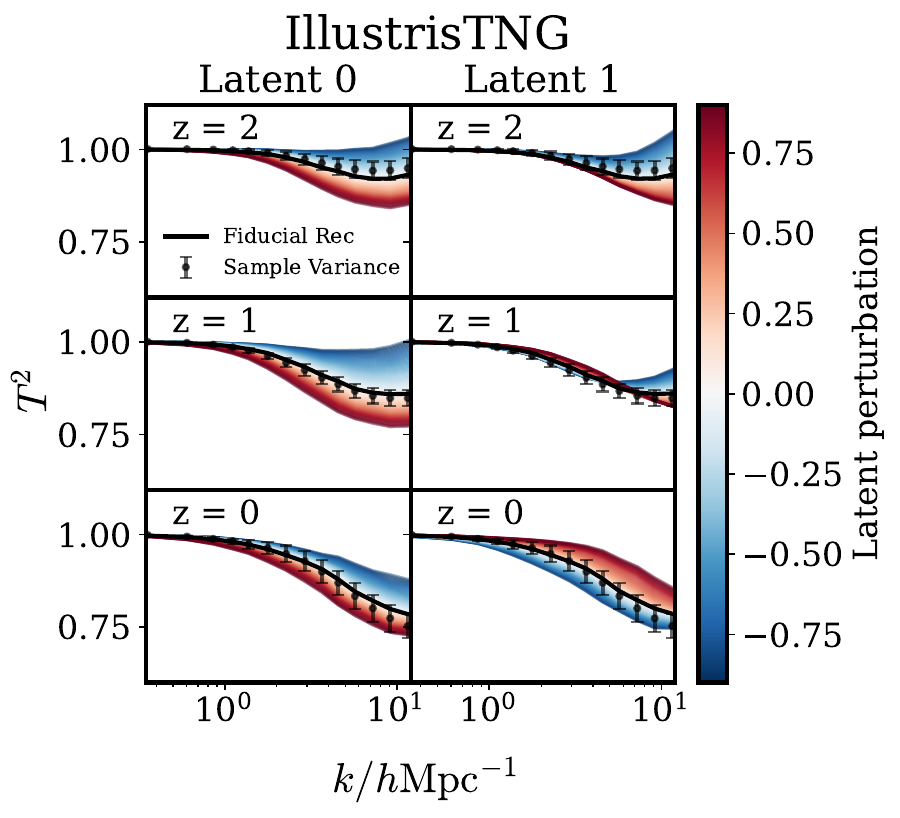}
  \includegraphics[height=0.4\textwidth]{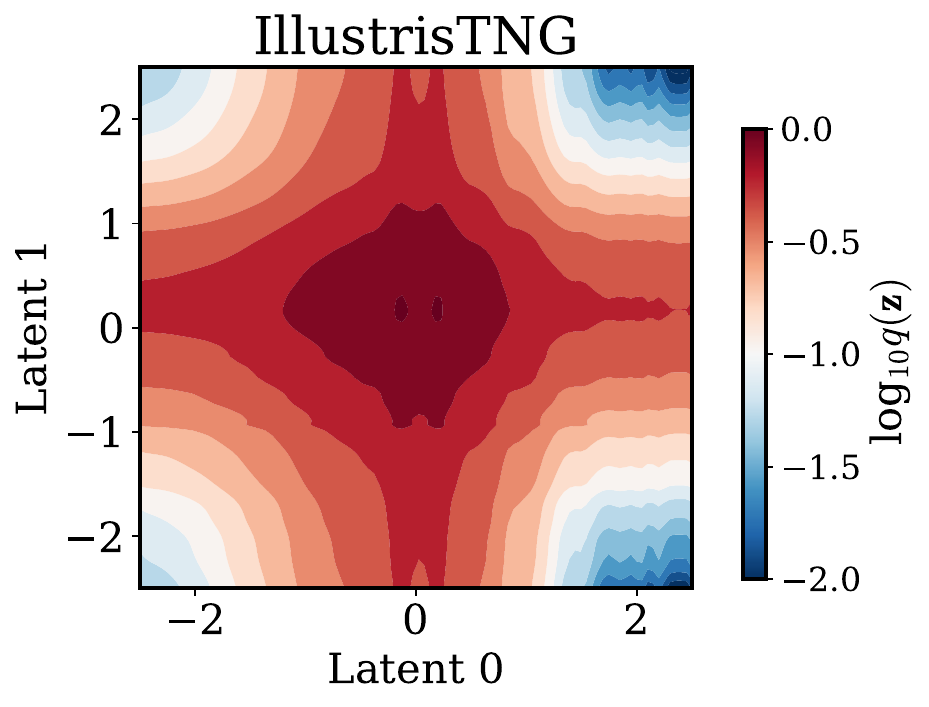}
  \includegraphics[width=\textwidth]
  {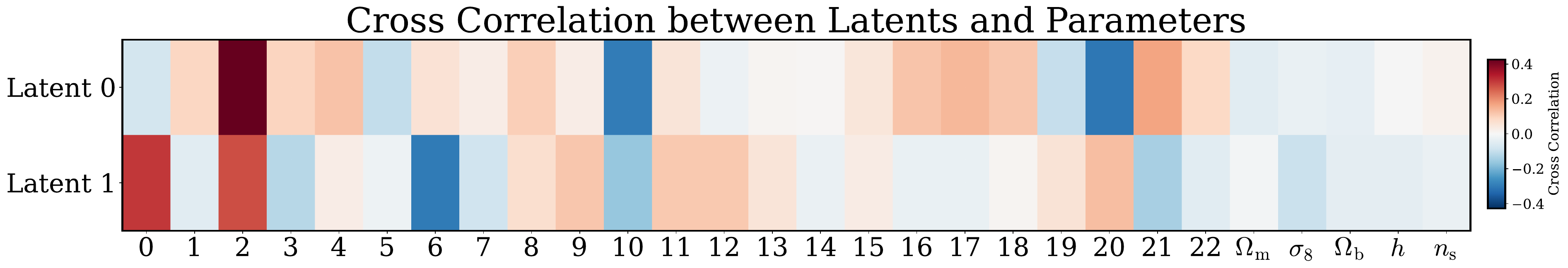}
  \caption{
    \textbf{Top}:
    Architecture of the conditional $\beta$-TCVAE.
    The encoder takes the baryonic feedback transfer function $\ln T^2$ and five cosmological parameters
    to infer the latent distribution and latent sample $\vz$.
    Latents and cosmology are passed to the decoder
    to reconstruct $\ln {T'}^2$.
    The architecture of the encoder and decoder can be found in \autoref{sec:en/de-coder}.\\
    \textbf{Mid left}:
    Effect of latent dimensions on $T^2$ reconstruction at $z=2$, $1$, and $0$ in IllustrisTNG.
    Black points and error bars show CV-set means and variances.
    Solid lines show reconstruction at the latent mean,
    with shading for latent deviation.\\
    \textbf{Mid right}:
    Latent PDF contours for the TEA model on IllustrisTNG,
    indicating disentanglement
    with a smooth distribution nearly factorizable along the principal axes.\\
    \textbf{Bottom}:
    Pearson cross-correlation between latents and parameters of IllustrisTNG SB28 simulation,
    computed from mean latents and parameters in the validation set.
    Colors indicate correlation strength.
    Parameters are listed in \autoref{Table:TNG_paras}.
     }
  \label{fig:AIO_plot}
\end{figure*}

\begin{figure*}[tb]
  \centering
  \includegraphics[width=0.8\textwidth]{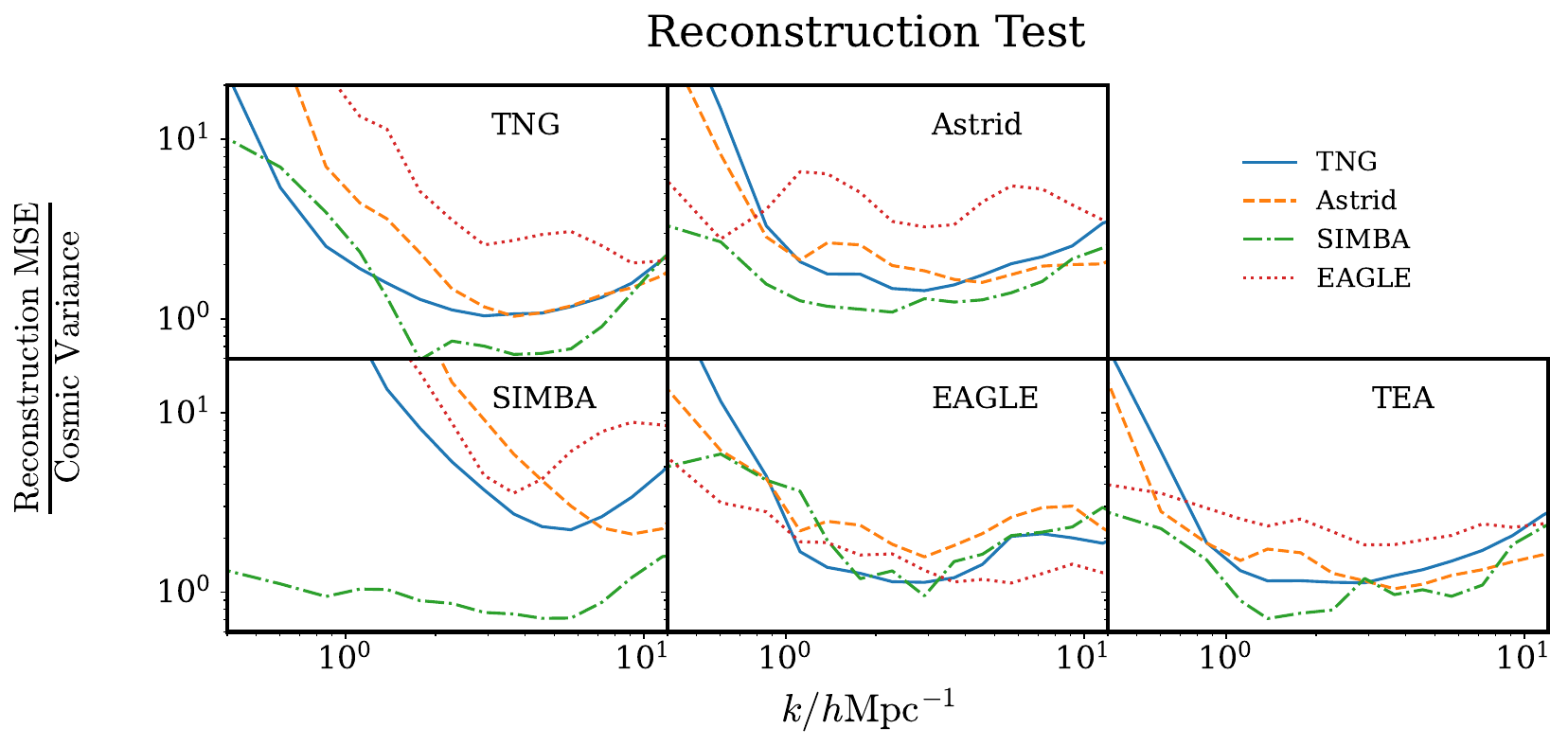}
  \caption{Reconstruction test showing $\frac{\mathrm{Reconstruction\ MSE}}{\mathrm{Cosmic\ Variance}}$ of the transfer functions $T^2$
  as a function of wavenumber, averaged over redshift.
  For each simulation in the CV set, 100 distinct
    reconstructions are created.
    The mean square error is calculated
    by averaging over the whole $27 \times 100 = 2700$ reconstructions.
    The redshift-averaged ratio was taken for each curve.
  Values near 1 indicate that reconstruction scatter is dominated by cosmic variance,
  with minimal additional model uncertainty.
  Subplots correspond to models trained on different datasets (\autoref{sec:training}).
  Colored lines represent test sets: IllustrisTNG (blue), Astrid (orange), SIMBA (green), and EAGLE (red).
  Each model achieves high reconstruction accuracy on its training suite.
  }
  \label{fig:chi2_test}
\end{figure*}

To verify that our model captures baryonic effects on $T^2$,
we first assess its reconstruction performance.
As we detail in Appendix~\ref{sec:recon_loss}, the reconstructions produced
by our model exhibit inherent uncertainty quantified by the latent posterior.
Concurrently, our dataset also features intrinsic scatter attributed
to cosmic variance.
To assess if the model's reconstruction can meet the noise levels
associated with cosmic variance, we utilize the CV dataset from CAMELS
introduced in \autoref{sec:data}.
The reconstruction of our model
aligns closely with the CV within the one-sigma range
as shown in the middle left panel of \autoref{fig:AIO_plot} .

We further quantify this by the ratio of reconstruction MSE to cosmic variance:

\begin{equation}
\frac{\mathrm{Reconstruction\ MSE}}{\mathrm{Cosmic\ Variance}} =
\frac{\mathrm{MSE}\left[T'^2(k, z|T^2_\textsc{cv}), T^2_\textsc{cv}(k, z)\right]}
{\mathrm{Var}\left[ T^2_\textsc{cv}(k, z) \right]},
\end{equation}
where $T^2_\textsc{cv}(k, z)$ represents the transfer functions of CV set,
and $T'^2(k, z \mid T^2_\textsc{cv})$
are reconstructions of the model using all
$T^2$ of the CV set as input.
This metric measures reconstruction uncertainty owing to the latent scatter
relative to the cosmic variance.
In the ideal case that VAE learns the baryonic representation perfectly,
the reconstruction is still subject to cosmic variance.
Therefore, this ratio should roughly be greater than 1,
and is smaller with better reconstruction.

The outcomes are shown in \autoref{fig:chi2_test}
for the five models trained across various datasets.
The ratio is elevated on large scales due to tiny
cosmic variance down to $10^{-6}$ in that range, and progressively
increases towards the small-scale end as short noise intensifies.
Each model performs well on its training suite,
though cross-suite performance varies.
The SIMBA-trained model performs well only
on SIMBA due to its strong feedback.

\subsection{Latent space structure}
\label{sec:latent}

With the reconstruction performance established in \autoref{sec:rec},
we now examine the latent representation.
The mid-right panel of \autoref{fig:AIO_plot} shows
the posterior $q(\vz) = \sum_{\vx_n} q(\vz|\vx_n)$
for the TEA model on IllustrisTNG,
revealing a smooth distribution with the two latent dimensions clearly disentangled.

This disentanglement is also evident in the reconstruction.
In the mid-left panel of \autoref{fig:AIO_plot},
we take $\overline{T^2_\textsc{cv}}$ as input,
extract the mean and standard deviation of each latent distribution,
and vary one latent dimension within $[\mu_i - 0.9, \mu_i + 0.9]$
while fixing the other at its mean.
The resulting reconstructions from the decoder are shown as colored bands.

As shown in the plot, ``Latent 0'' enhances the spectrum suppression
in the same way across all redshifts, and tends to induce feedback on large-scale.
But the effect of ``Latent 1'' evolves with time,
as a large ``Latent 1'' would suppress the power spectrum more strongly
at $z=2$ but diminish that suppression at $z=0$.
This on one hand provides additional evidence for the disentanglement
of the latent space.
On the other hand, the presence of two separate patterns within the
latent dimensions provides hints for the physics behind them.

\subsection{Correlation between latent and simulation parameters}
\label{sec:cc}

\begin{figure*}[tb]
\centering
\includegraphics[width=\textwidth]{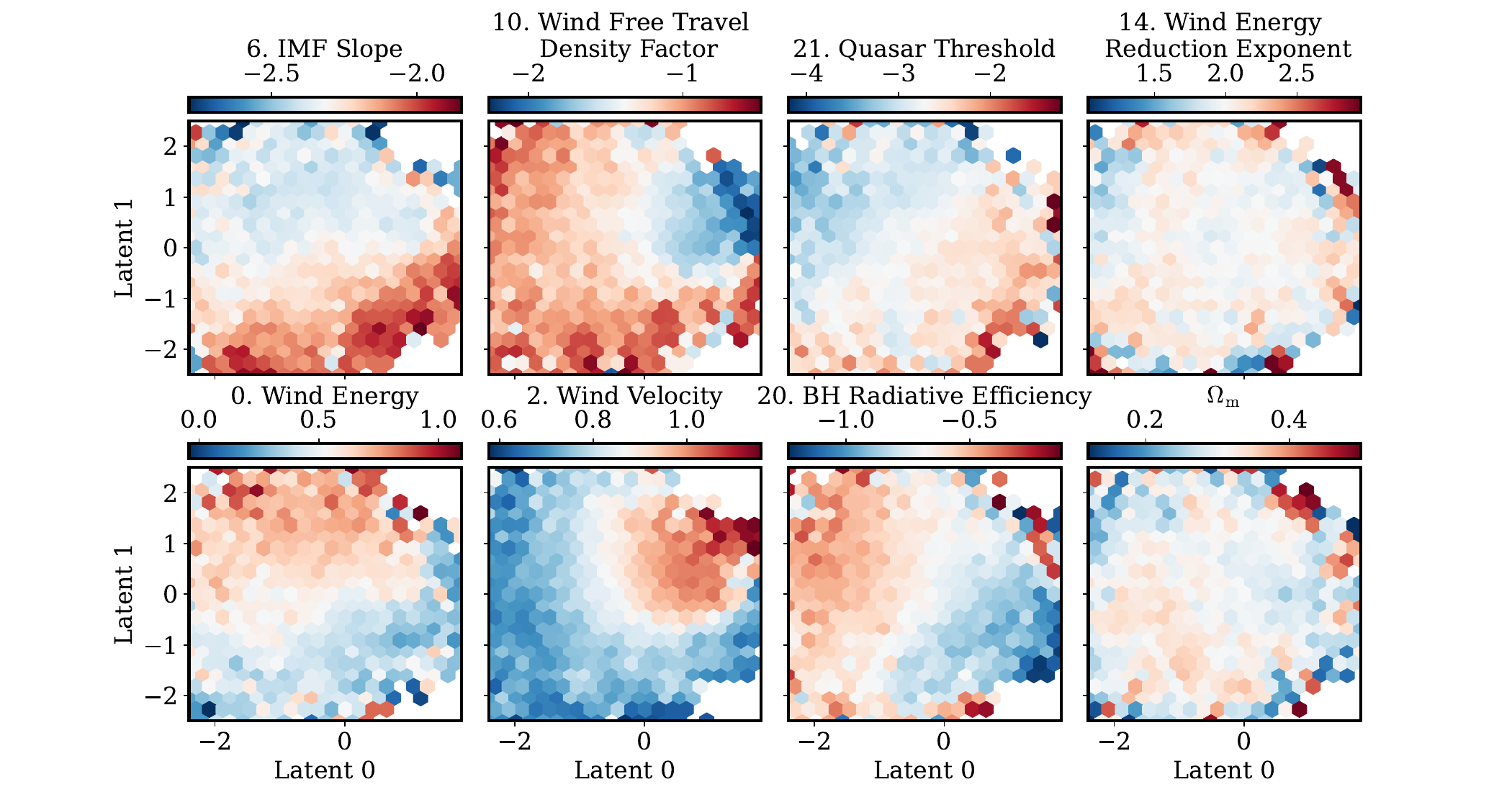}
\caption{
  Heatmaps of simulation parameter distributions projected onto the latent space.
  For each simulation in the IllustrisTNG suite,
  we generate 20 latent samples and compute the mean value of each parameter within 2D hexagonal bins.
  Each panel shows the average value of a specific parameter across the latent space.\\
  From left to right, the first six panels display parameters
  with the strongest cross-correlations with the latent dimensions,
  all showing structured and consistent patterns,
  including detailed nonlinear features.
  In the right column, we present two representative parameters (a baryonic one and $\Omega_\mathrm{m}$) that exhibit negligible correlation with the latent space.
  Their heatmaps appear nearly uniform, consistent with their low cross-correlation values.}

\label{fig:contour_TNG}
\end{figure*}

We examine how the latent parameters capture baryonic feedback
by computing cross-correlations between the two latents
and the 28 parameters of the IllustrisTNG SB28 simulations.
For each simulation, we take the parameter values $\alpha_i$
and the mean of each latent's Gaussian posterior from the encoder,
yielding $28\times2$ correlations, shown in \autoref{fig:AIO_plot}.
Both latents show negligible correlation with cosmological parameters
but distinct patterns with baryonic physics parameters,
consistent with the disentanglement in \autoref{sec:latent}.

Of the 23 baryonic parameters, six have strong correlations (Pearson's coefficients larger than $0.15$):
(0) Wind Energy, (2) Wind Speed, (6) IMF Slope, (10) Density of Wind Recoupling,
(20) BH Radiative Efficiency, and (21) Quasar Threshold.
Their distributions in latent space are shown in \autoref{fig:contour_TNG},
with good agreement with the correlations in \autoref{fig:AIO_plot}.
These parameters form three pairs with opposite correlation signs within each pair,
suggesting complementary feedback roles:
\begin{itemize}
    \item \textbf{(0) Wind Energy-(6) IMF slope:}
    Both mainly correlate with ``Latent 1''.
    High (0) Wind Energy strengthens the SN-driven winds, enhancing supernova feedback,
    while a steeper (6) IMF Slope has an opposite effect
    by increasing the metallicity gas cooling \citep{Lee2024feedback}.
    This results in a strong positive correlation between SN feedback and ``Latent 1''.
    In addition, black hole (BH) mass is also affected by (0) Wind Energy (negatively) and (6) IMF Slope (positively).
    Thus, ``Latent 1'' shows a negative correlation with BH feedback.

    \item \textbf{(2) Wind Speed – (10) Wind Free Travel Density Factor:}
    Both affect wind propagation range and correlate with both latents,
    more strongly with ``Latent 0'',
    influencing the scale of the feedback.

    \item \textbf{(20) BH Radiative Efficiency – (21) Quasar Threshold:}
    This pair strongly correlates with ``Latent 0'', and weakly with ``Latent 1'' in the opposite direction.
    Lower (20) BH Radiative Efficiency promotes BH growth,
    while a higher (21) Quasar Threshold
    favors kinetic feedback at lower accretion rates. \\
    As a result, larger ``Latent 0'' values imply stronger BH feedback,
    while ``Latent 1''  tends to suppress it.

\end{itemize}

\subsection{How latents affect the spectrum}
\label{sec:feedback}

\begin{figure*}[tb]
\centering
\includegraphics[width=0.8\textwidth]{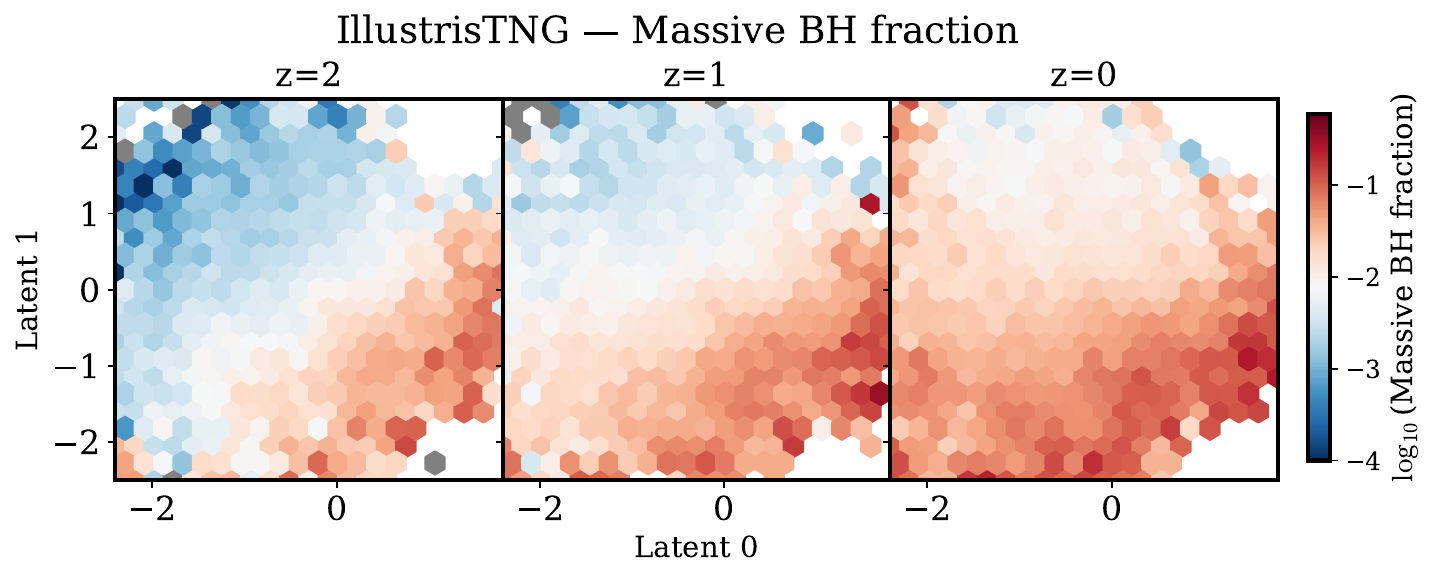}
\includegraphics[width=0.6\textwidth]{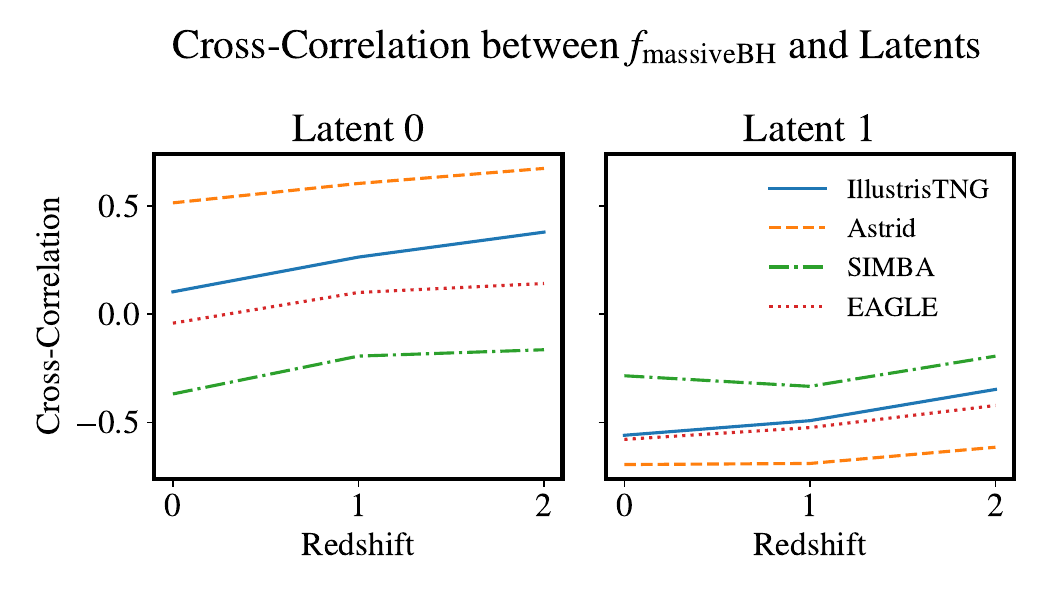}
\label{fig:mBH_cc}
\caption{
\textbf{Upper}:
Heatmaps of the massive black hole fraction,
$\log_{10}(f_\mathrm{massive\ BH})$ with $M_\mathrm{BH} \geq 10^8\,M_\odot$,
in the latent space across $z=2$, $1$, and $0$ (right to left)
for IllustrisTNG, similar to \autoref{fig:contour_TNG}.
At all redshifts, ``Latent 0'' correlates positively with BH mass,
while ``Latent 1'' correlates negatively.\\
\textbf{Lower}:
Cross-correlation between latent variables and the massive black hole fraction
across redshift for all four simulation suites.
``Latent 0'' shows positive correlations for IllustrisTNG and Astrid, increasing with redshift,
but a negative trend for SIMBA.
Negative correlations are shown with ``Latent 1'' for all suites.
}

\end{figure*}

In this section, we use the correlation patterns in \autoref{sec:cc}
to interpret the latent effects on $T^2$ reconstruction (\autoref{sec:latent})
and uncover their physical meaning.

Based on \autoref{sec:cc}, ``Latent 0'' is positively correlated
with the BH feedback via (20) BH Radiative Efficiency – (21) Quasar Threshold,
while a large ``Latent 1'' suppresses the BH feedback
through combined effects from (0) Wind Energy, (6) IMF Slope,
(20) BH Radiative Efficiency, and (21) Quasar Threshold.
To explore this, we examine the fraction of massive black holes
($M_\mathrm{BH} \geq 10^8 M_\odot$, denoted as $f_\mathrm{massive\ BH}$),
across different redshifts in the IllustrisTNG suite,
finding consistent trends between BH mass and the latents across all redshifts, as shown in the upper panel of \autoref{fig:mBH_cc}.

Since stronger black hole feedback suppresses the power spectrum,
a large ``Latent0'' would enhance suppression,
while a large ``Latent 1'' should weaken it.
This is clear at $z=0$ (\autoref{fig:AIO_plot}, mid-left).
But at $z=2$, both latents drive suppression,
implying other processes beyond BH feedback are acting at early times.

SN feedback explains this discrepancy.
Correlation between (0) Wind Energy–(6) IMF Slope strongly and ``Latent 1''
introduces SN-driven suppression,
explaining the suppression at $z=2$.
Meanwhile, (2) Wind Speed - (10) Wind Free Travel Density Factor correlates with ``Latent 0'',
yielding stronger large-scale suppression than ``Latent 1''.

To test these hypotheses, we examine two special simulations:
one with no BHs at all, and another with BHs but without the effective (kinetic) AGN feedback mode.
In both cases, AGN feedback is effectively absent, and SN feedback dominates.
We observe suppression of the power spectrum at $z = 2$,
supporting the idea that SN feedback, not AGN feedback,
is responsible for the suppression at $z=2$.
Also, similar latent values are assigned by our model:
$(-1.51 \pm 0.23, 1.51 \pm 0.63)$ for the no-BH case,
and $(-1.51 \pm 0.22, 1.81 \pm 0.63)$ for the no-kinetic-feedback case.
This matches our earlier interpretation:
negative Latent 0 and huge Latent 1 reflect reduced BH feedback,
with big ``Latent 1'' coinciding with the leading SN feedback at the same time.

In summary, a large ``Latent 0'' corresponds to increased BH growth and stronger suppression of the power spectrum across all redshifts, especially at large scales due to farther-traveling winds.
In contrast, ``Latent 1'' negatively correlates with BH mass and positively correlates with SN feedback,
resulting in suppression at $z = 2$ and
diminished suppression or even enhancement at $z = 0$ for large ``Latent 1'' values.
These trends are consistent across different simulation suites,
as shown by the correlation between latents and BH mass in \autoref{fig:mBH_cc}.

\subsection{Stability of latent space dimension}
\label{sec:3d_model}

\begin{figure*}[tb]
\centering
\includegraphics[height=0.35\textwidth]{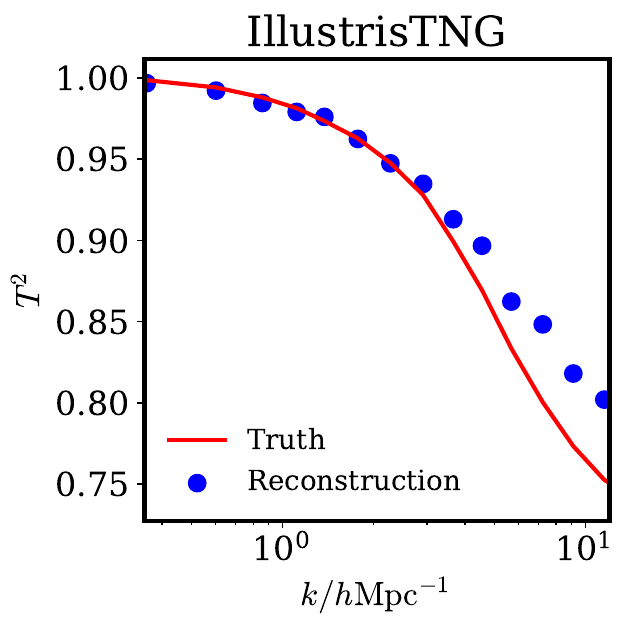}
\includegraphics[height=0.35\textwidth]{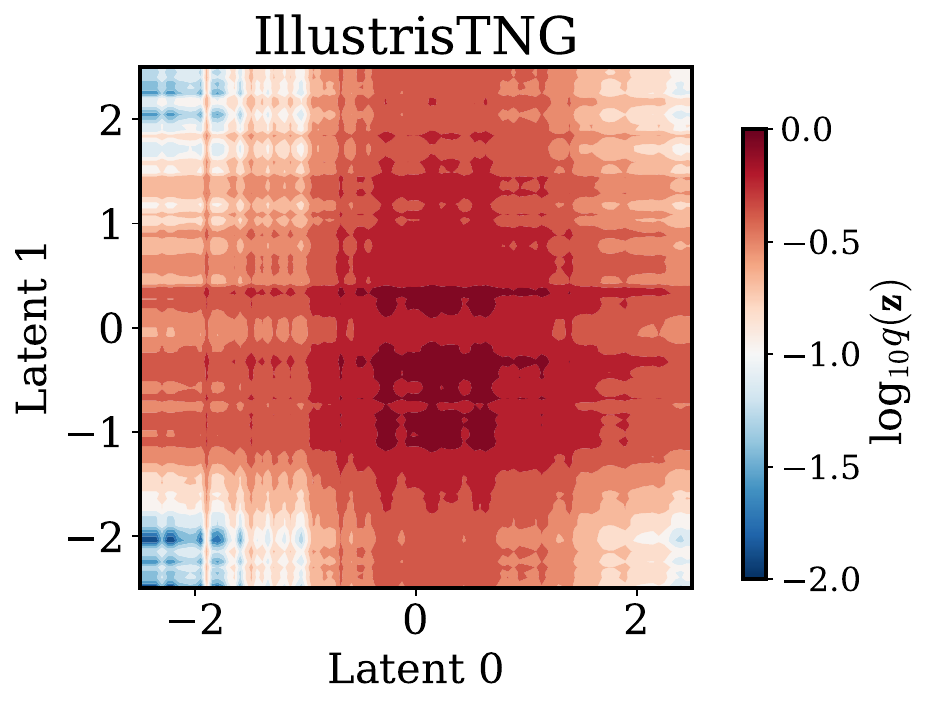}
\includegraphics[width=0.7\textwidth]{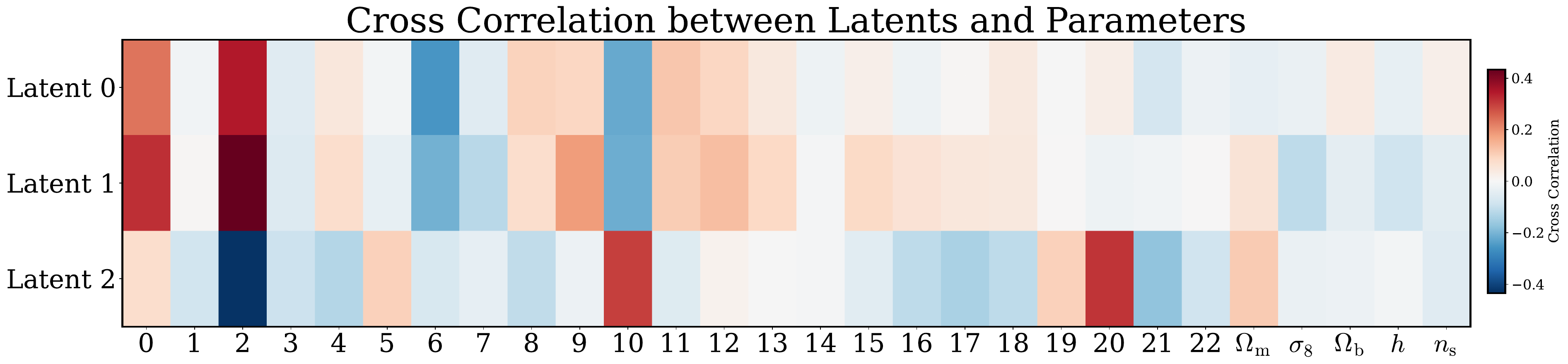}
\caption{
    Examples of three types of failure behavior observed in 3D latent space models.\\
    \textbf{Top left:}
    Incomplete reconstruction from a 3D model.
    The red line shows the mean power spectrum ratio from the CV set,
    while the blue dots indicate the corresponding reconstructed spectra ratio.
    The reconstruction significantly deviates
    from the truth for $k > 1\, h\,\mathrm{Mpc}^{-1}$.\\
    \textbf{Top right:}
    An example of an overly concentrated latent distribution.
    The sharply striped contour indicates that
     at least one of the latents has collapsed into a nearly delta-function-like distribution.\\
    \textbf{Bottom:}
    Cross-correlation between the three latent dimensions
    and simulation parameters similar to the bottom panel of \autoref{fig:AIO_plot}.
    For visual clarity, the sign of ``Latent 1'' has been flipped in this specific example
    to highlight its similarity to ``Latent 0''.
}
\label{fig:bad_3d}
\end{figure*}

Even though our model with a two-dimensional latent space works quite well
on the general datasets, we further test whether a higher-dimensional latent space
would lead to better performance.
We first performed a principal component analysis (PCA) for the transfer functions.
For all suites, the explained variance fractions
for the first three components are $70 \sim 80\%$, $10 \sim 15\%$, and $<5\%$,
respectively, indicating that only the first two components are significant,
which is consistent with early PCA results \citep{Huang2019bar}.
As PCA is computed directly with $T^2$, it contains cosmology.
Thus, the number of principal components
should be no less than our latent dimensionality,
supporting our 2D latent space.
As for the reconstruction comparison between PCA and VAE, more details can be found in \autoref{app:PCA}.

To confirm that two latent dimensions suffice,
we further train 3D models using the same strategy as in \autoref{sec:training}.
However, the models on the Pareto front exhibit three types of misbehavior
as shown in \autoref{fig:bad_3d}: (1) incomplete reconstruction, in the top left panel,
(2) an oddly behaving latent space
(either overly dispersed or highly concentrated) in the top right panel,
and (3) a failure to disentangle, which is shown in the bottom panel,
where two latent dimensions are strongly correlated
or there is one meaningless latent.
The first two indicate overfitting,
while the third suggests the model attempts to
replicate an existing 2D latent direction rather than learning a new one.
This demonstrates the model's inability to form an independent third latent.

Combining these results with the physical interpretation
of the 2D latent space in \autoref{sec:feedback},
we conclude that two latent dimensions
are sufficient to capture baryonic physics
across the four CAMELS suites.

\section{Discussion}
\label{sec:Discussion}

In this paper, we developed a probabilistic machine learning model
based on the $\beta$-TCVAE architecture to learn
a general, low-dimensional latent representation of baryonic effect on matter power spectrum
from multiple CAMELS simulation suites.
The resulting \textbf{two-dimensional} latent space is both disentangled
and minimally correlated with cosmological parameters,
enabling simulations from different suites to be coherently mapped
into a unified representation.
Previous methods for modeling baryonic physics
often struggle to generalize across suites and
suffer degraded reconstruction accuracy from degenerate parameterizations
\citep{zhou2025emulator, Schaller2025anl_bar, Schaller2025Flamingo, kammerer2025sr}.
But our unified latent space supports seamless generalization
and provides a foundation for future simulation-based inference.
Furthermore, we find that the two latent dimensions modulate the transfer function $T^2$
primarily related to BH growth and SN feedback within the simulations we examined.

At present, our architecture shows biases when SIMBA is included in the training set,
likely due to its extreme feedback model, and our physical interpretation remains largely driven by IllustrisTNG.
These limitations could be alleviated with the upcoming second-generation CAMELS simulations,
which feature larger volumes and more sophisticated baryonic physics.
However, as shown in \autoref{fig:chi2_test},
the TEA model can still reach a decent reconstruction on SIMBA within the range expected from cosmic variance.
On the other hand, though we already include over $5,000$ simulations from 4 different suites,
a broader and more diverse dataset would enable the model to capture more complex and extreme feedback phenomena.
In parallel, more advanced architectural extensions could be explored once supported by richer training data.

Given that the two latent dimensions correlate strongly with only six baryonic parameters,
it may be possible to derive an analytical mapping with symbolic regression, offering qualitative insight into feedback mechanisms.
The decoder also serves as a fast emulator, rescaling gravity-only matter power spectra into their hydrodynamic counterparts.
A similar approach could be extended to other summary statistics or observables, such as weak-lensing $3\times2$pt correlation functions.

Looking ahead, the upcoming Stage-4 surveys, such as LSST, Euclid, and Roman
will demand accurate modeling of baryonic effects in weak lensing analysis.
With the transfer functions generated by our model, baryonic contributions can be separated from cosmology, thereby extending cosmological studies to smaller scales.
By encoding spectra from real observations,
our framework could directly compare feedback models
against data through weak-lensing correlation functions,
calibrating the latent space to real data
while enabling systematic comparisons between simulations and observations.

\begin{acknowledgements}

We gratefully acknowledge the valuable comments provided by Xin Liu.
Y.L.\ is supported by the Major Key Project of Peng Cheng Laboratory and
the National Key Research and Development Program of China under grant
number 2023YFA1605600.
S.L.\ acknowledges support by Illinois Campus Research Board Award
RB25035 and NSF grant AST-2308174.
The Flatiron Institute is supported by the Simons Foundation.
B.D.\ acknowledges support from the Ambrose Monell Foundation, the
Corning Glass Works Foundation Fellowship Fund, and the Institute for Advanced Study.
This work utilizes resources supported by the National Science Foundation’s Major Research Instrumentation program, grant \#1725729, as well as the University of Illinois at Urbana-Champaign \citep{HAL}.

\end{acknowledgements}

%

\software{\texttt{PyTorch} \citep{PyTorch}, \texttt{Optuna}
\citep{Optuna}, \texttt{NumPy} \citep{NumPy}, \& \texttt{matplotlib}
\citep{matplotlib}.}

\FloatBarrier
\appendix

\section{Comparison between Methods}
\label{sec:compare_methods}

\begin{table}[]
\centering
\caption{Comparison across five representative methods:
baryonic correction model (BCM), symbolic regression (SR), physically motivated parameterizations (PMP), principal component analysis (PCA), and our $\beta$-TCVAE method.
$\checkmark$ = yes and $\times$ = no.
D means data-driven, which typically uncover more parsimonious
representations, and M stands for hand-crafted, that could result in
large parameterizations.}
\newcommand{\y}{\checkmark}
\newcommand{\n}{\(\times\)}
\renewcommand{\arraystretch}{1.15}

\begin{tabular}{lccccc}
\toprule
Criterion & BCM & SR & PMP & PCA & VAE (this work) \\
\midrule
Driven by Data/Modeler       &  M &  D &  M &  D &  D \\
Subgrid Model Independence   & \y & \n & \y & \y & \y \\
Cosmology Independence       & \y & \y & \n & \n & \y \\
Redshift Independence        & \y & \y & \n & \y & \y \\
Interpretability             & \y & \n & \y & \n & \y \\
Disentangled Representation  & \n & \n & \n & \n & \y \\
\bottomrule
\end{tabular}
\label{table:compare_methods}
\end{table}

{
As discussed in \autoref{sec:intro}, numerous studies have investigated baryonic feedback effects using the transfer function.
Here, we summarize the advantages and limitations of five methods:
baryonic correction models (BCM), symbolic regression (SR), physically motivated parameterizations (PMP), principal component analysis (PCA), and our $\beta$ total correlation variational autoencoders ($\beta$-TCVAE) in \autoref{table:compare_methods}.
}

\section{CAMELS simulations}
\label{sec:camels}

As introduced in \autoref{sec:data}, we use four suites from CAMELS:
IllustrisTNG SB28, Astrid SB7, SIMBA LH6, and Swift-EAGLE LH6.
Here we summarized the basic properties of the four suites in \autoref{table:CAMELS}.

Especially, the IllustrisTNG SB28 suite employs a 28-dimensional Sobol sequence to uniformly sample cosmological and astrophysical parameter space, providing broad and efficient coverage of high-dimensional variations.
Each simulation utilizes a unique parameter combination and random seed, allowing for both systematic parameter studies and assessments of cosmic variance.
Runs begin at $z=127$ with 2LPT initial conditions, include matched dark-matter-only counterparts, and output 91 snapshots (including $z=0, 1, 2$ that we use), with halos and merger trees identified using multiple standard finders.
We summarize the 28 parameters in \autoref{Table:TNG_paras}.

\begin{table*}[tb]
\centering
\caption{Summary of the four CAMELS simulation suites used.}
\label{table:CAMELS}
\begin{tabular}{lccccr}
\toprule
\textbf{Suite} & \textbf{Number} & \textbf{Code} & \textbf{Subgrid Model} & \textbf{Dimension} & \textbf{Sampling} \\
\midrule
IllustrisTNG SB28 & 2,048 & AREPO & IllustrisTNG & 28D & Sobol sequence \\
Astrid SB7        & 1,024 & MP-Gadget & ASTRID   & 7D & Sobol sequence  \\
SIMBA LH6         & 1,000 & GIZMO & SIMBA        & 6D & Latin Hypercube \\
Swift-EAGLE LH6   & 1,000 & Swift & EAGLE        & 6D & Latin Hypercube \\
\bottomrule
\end{tabular}
\end{table*}

\begin{table*}[htbp]
\centering
\caption{Description of parameters of IllustrisTNG SB28}
\label{Table:TNG_paras}
\begin{tabular}{rlrlrl}
\toprule
\textbf{Idx} & \textbf{Parameter} & \textbf{Idx} & \textbf{Parameter} \\
\midrule
0  & Wind Energy                  & 1  & Radio Feedback Factor \\
2  & Wind Speed                   & 3  & Radio Feedback Reorientation \\
4  & Max SFR Timescale            & 5  & Factor for Softer EOS \\
6  & IMF Slope                    & 7  & SNII Min Mass \\
8  & Thermal Wind Fraction        & 9  & Variable Wind Spec Momentum \\
10 & Wind Free Travel Density Factor & 11 & Min Wind Speed \\
12 & Wind Energy Reduction Factor & 13 & Wind Energy Reduction Metallicity \\
14 & Wind Energy Reduction Exponent & 15 & Wind Dump Factor \\
16 & Seed Black Hole Mass         & 17 & Black Hole Accretion Factor \\
18 & Black Hole Eddington Factor  & 19 & Black Hole Feedback Factor \\
20 & Black Hole Radiative Efficiency & 21 & Quasar Threshold \\
22 & Quasar Threshold Power       & 23--27 & Cosmological Parameters \\
\bottomrule
\end{tabular}
\end{table*}

\section{Deduction of the ELBO of VAE}
\label{sec:VAE}

\subsection{The family of VAE}
\label{sec:VAEs}

Variational autoencoders (VAE) are a latent variable model consisting of
three parts: an ``Encoder'', a ``Decoder'' and a ``Bottleneck'' in
between \citep{kingma2013autoencoding}.
For a given input sample, the encoder will map the sample $\vx$ to an
element $\vz$ in the bottleneck, called ``latent'' in our case, following
a posterior distribution corresponds to $\vx$, $q(\vz|\vx)$.
With the decoder, we could generate a new sample for a given latent,
with the distribution $p(\vx'|\vz)$.

VAE is trained by optimizing the tractable evidence lower bound (ELBO):
\begin{align}
  \Loss_\mathrm{VAE} = \frac{1}{N}
  \sum_{n=1}^{N} (\mathbb{E}_{q(\vz|\vx_n)}[\log p(\vx_n|\vz)]-\KL(q(\vz|\vx_n)||p(\vz))).
\label{eq:VAE_ELBO}
\end{align}

In the first term, $p(\vx_n|\vz)$ stands for
the likelihood of a certain sample $\vx_n$ given the latent $\vz$.
If the $\vz$ with a high value of $q(\vz|\vx_n)$ (i.e., the latent value $\vz$
preferred by the sample $\vx_n$) also has a large $p(\vx_n|\vz)$ value (i.e.,
likely to reconstruct the data), the first term would be large.
So, the first term indicates the reconstruction power of VAE and is
usually called ``Reconstruction Loss''.

In the second term, $p(\vz)$ stands for the prior distribution of the
latent, a standard Gaussian distribution is our case.
When this term is small, the posterior closely matches the prior,
implying that the latent representation carries little information about the input. In the extreme case where the KL divergence vanishes,
the latent space has zero information capacity.
Therefore, a larger KL term is often desirable to ensure informative latent encodings—this is why the ELBO is maximized despite the negative sign in front of the KL divergence term.
In the extreme case where the KL divergence vanishes, the latent space has zero information capacity. Therefore, a larger KL term is often desirable to ensure an informative latent.

In practice, we would like different dimensions of the latent to be
aligned with the components that have different contributions to
reconstruction.
So, there is $beta$-VAE, with the following ``modified'' ELBO:
\begin{align}
  \Loss_\beta = \frac{1}{N}
  \sum_{n=1}^{N} \left(\mathbb{E}_q[\log p(\vx_n|\vz)]-\beta * \KL(q(\vz|\vx_n)||p(\vz))\right).
\end{align}
With $\beta>1$, the effect of the second term would be enhanced so that
a more disentangled latent may be
achieved \citep{burgess2018understanding}.

\textbf{$\beta$-TCVAE}:
Even though $\beta$-VAE may provide us with more disentangled latent
representation, it may end up with a worse reconstruction, as much
information is lost when $q(\vz|\vx)$ approaches a standard
Gaussian.
To solve this problem, we may need
to dig into the KL term in the ELBO.
As shown in \citep{chen2019isolating}, that term can actually be
decomposed into three KL terms:
\begin{align}
  &\frac{1}{N}\sum_{n=1}^{N} (\KL(q(\vz|\vx_n)||p(\vz)))\\
  =&\mathbb{E}_{p(\vx_n)}[\KL(q(\vz|\vx_n)||p(\vz))]\\
  =&\mathbb{E}_{p(\vx_n)}[
    \mathbb{E}_{q(\vz|\vx_n)}[
      \log(q(\vz|\vx_n))-\log(p(\vz))+\\
      &\log(q(\vz))-\log(q(\vz))+\log(\prod_j q(\vz_j))-\log(\prod_j q(\vz_j))
    ]
  ]\\
  =&\mathbb{E}_{q(\vz,\vx_n)}[\log(\frac{q(\vz|x
  _n)}{q(\vz)})]
  +\mathbb{E}_{q(\vz)}[\log(\frac{q(\vz)}{\prod_j q(z_j)})]
  +\mathbb{E}_{q(\vz)}[\log(\frac{\prod_j q(z_j)}{p(\vz)})]\\
  =& \mathbb{E}_{q(\vz,\vx_n)}[\log(\frac{q(\vz,\vx_n)}{q(\vz)p(\vx_n)})]
  +\mathbb{E}_{q(\vz)}[\log(\frac{q(\vz)}{\prod_j q(z_j)})]
  +\sum_{j} \mathbb{E}_{q(z_j)}[\log(\frac{q(z_j)}{p(z_j)})]\\
  =& \KL(q(\vz,\vx_n)||q(\vz)p(\vx_n)) + \KL(q(\vz)||\prod_j q(z_j))
  + \sum_{j} \KL(q(z_j)||p(z_j)).
  \label{eq:KL_decomp}
\end{align}

These are the three KL term introduced in \autoref{sec:VAE_main}.
As we would like to find an explainable latent representation for the
baryonic feedback effects with the transfer function,
we choose $\beta$-TCVAE to seek a set of disentangled latents
that have desired information with feedback effects.

\subsection{Reconstruction Loss}
\label{sec:recon_loss}

For a common Gaussian decoder (i.e., the posterior $p(\vx_n|\vz)$ is a
standard Gaussian), the reconstruction loss can be simplified as the
Mean Squared Error (MSE) loss:
\begin{equation}
  \begin{aligned}
    \mathbb{E}_{q(\vz|\vx_n)p(\vx_n)}[\log p(\vx_n|\vz)] &= \frac{1}{2} || \vx' - \vx ||^2 + \log\sqrt{2\pi}\\
    &= \frac{D}{2} \mathrm{MSE}(\vx' - \vx) + c,
  \end{aligned}
  \label{eq:recon}
\end{equation}

where $D$ is the dimensionality of the input $\vx$, and $\vx'$ denotes its reconstruction.
However, using a simple MSE loss can be limiting as it assumes an
identity covariance matrix, which may not capture the complexity of the
data distribution effectively.

To improve reconstruction fidelity, we adopt a more expressive \emph{variational decoder}, where the likelihood $p(\vx_n|\vz)$ is modeled as a Gaussian with a non-identity, diagonal covariance matrix:
\begin{align}
    p(\vx_n|\vz) = \mathcal{N}(\mu_\theta(\vz), \sigma_\theta(\vz)^2).
\end{align}
Each dimension $i$ of the data (corresponding to different redshifts and wavevectors $k$ of the input spectrum ratio) is assigned a unique variance parameter. Following the $\sigma$-VAE design proposed in~\cite{Rybkin2021simple}, we set the variance based on the input and reconstruction:

\begin{equation}
  \sigma_{\theta, i}^2 = \mathrm{MSE}(\vx'_i, \vx_i),
\end{equation}

and the reconstruction loss becomes:
\begin{equation}
    \mathbb{E}_{q(\vz|\vx_n)p(\vx_n)}[\log p(\vx_n|\vz)] =
    -\sum_{i} \left( \log \sigma_{\theta, i} + \frac{1}{2 \sigma_{\theta, i}^2} \, \mathrm{MSE}(\vx'_i, \vx_i) \right)
    \label{eq:rec_loss}
\end{equation}

When training, gradient flow through $\sigma_{\theta, i}$ is stopped during backpropagation. As a result, only $\mathrm{MSE}(\vx'_i, \vx_i)$ in the second term contribute gradient, and the variance we add would only serve as a weight. This formulation enables the model to adaptively weigh reconstruction errors, allowing for a more flexible and accurate representation of the data.

Combining the results from \autoref{eq:KL_decomp} and \autoref{eq:rec_loss}, the ELBO for $\beta$-TCVAE can be written as:
\begin{align}
\Loss_{\beta-\mathrm{TC}} =
&-\sum_{i} \Bigl[ \log \sigma_{\theta, i} + \frac{1}{2 \sigma_{\theta, i}^2} \, \mathrm{MSE}(\vx'_i, \vx_i) \Bigr] \nonumber\\
&- [\alpha \KL\bigl( q(\vz,\vx_n) \mVert q(\vz)p(\vx_n) \bigr) \nonumber\\
&+ \beta \KL\bigl( q(\vz) \mVert \tprod_j q(z_j) \bigr) \nonumber\\
&+ \gamma \sum_j \KL\bigl( q(z_j) \mVert p(z_j) \bigr)].
\label{loss:beta-TC}
\end{align}
\autoref{loss:beta-TC} generalizes VAE, which is a special case when
$\alpha = \beta = \gamma = 1$.
And if $\alpha = \beta = \gamma > 1$, it becomes $\beta$-VAE.
Appendix~\ref{sec:cal_loss} details the loss computation methods.

\section{Calculation of the Losses}
\label{sec:cal_loss}

\subsection{Sampling in Minibatch}
\label{sec:minibatch}

When training and tuning the model, we usually deal with a minibatch
instead of the whole dataset.
So, we would need a method to estimate the value of $q(\vz)$ with
minibatch $\batch_M$.
With the ``Minibatch Stratified Sampling'' (MSS) method \citep{chen2019isolating}, it has been shown
that an unbiased estimator exists:
\begin{align}
  f(\vz, n*, \batch_M) = \frac{1}{N}q(\vz|\vx_{n*})+\frac{1}{M-1}\sum_{m=1}^{M-2}q(\vz|\vx_m)+\frac{N-(M-1)}{N(M-1)}q(\vz|\vx_{M-1}),
\end{align}
where $M$ is the batch size and $\vz$ is originally sampled from
$q(\vz|\vx_{n*})$.
For any given $\vz$ and $n*$, if we sum over all the possible
combination of $\vx_{M-1}$ and take the average, then each
$q(\vz|\vx_n)$ would appear in the sum
$\frac{1}{M-1}\sum_{m=1}^{M-2}q(\vz|\vx_m)$ for $(M-2)$ times and appear
as $\frac{N-(M-1)}{N(M-1)}q(\vz|\vx_{i})$ once.
So, for each $\vx_n$, the total contribution would be:
\begin{equation}
  \begin{aligned}
    & \frac{M-2}{M-1} q(\vz|\vx_n) + \frac{N-(M-1)}{N(M-1)} q(\vz|\vx_n) \\
    & = \frac{N(M-2)+N-(M-1)}{N(M-1)}q(\vz|\vx_n) \\
    & = \frac{N(M-1)-(M-1)}{N(M-1)}q(\vz|\vx_n) \\
    & = \frac{N-1}{N}q(\vz|\vx_n).
  \end{aligned}
\end{equation}

Averaging over \{$\vx_{1}$...$\vx_{(M-1)}$\}, we would have a new
estimator:
\begin{align}
  \bar{f}(\vz, n*, \batch_M) = \frac{1}{N}q(\vz|\vx_{n*})+\frac{N-1}{N(M-1)}\sum_{m=1}^{M-1}q(\vz|\vx_m).
  \label{eq:q_est}
\end{align}

It can be easily seen that \autoref{eq:q_est} returns to the definition of
$q(\vz)$ (i.e., $q(\vz) = \frac{1}{N}\sum_n q(\vz|\vx_n)$) when $M = N$.

\subsection{Calculation of KL terms}
\label{sec:KL_loss}

For each $\vx_n$ from the minibatch $\hat{B}_M$, we denote the
corresponding latent sample as $\vz_n$, which follows $q(z|x)$.
We label the minibatch of $\vz_n$ as $q(\batch_M)$.
$q(\batch_M)$ can also be equivalently considered as sampling by
$q(\vz)$ from the whole parameter space.
For any $\vz_n$, we can estimate the value of $q(\vz_n)$ by the
estimator from the last section.

So the value of MI-loss can be estimated by:
\begin{equation}
\begin{aligned}
  \Loss_\mathrm{MI} &=
  \mathbb{E}_{q(\vz,\vx_n)}[\log(\frac{q(\vz|\vx)}{q(\vz)})] \\
  & = \overline{\ln q(\vz_n|\vx_n)-\ln q(\vz_n)}
\end{aligned}
\end{equation}

Here, the overline indicates averaging over the whole $q(\batch_M)$.
As $q(\batch_M)$ is constructed by sampling with $q(\vz|\vx)$, averaging
over it can be considered an estimator of the expectation.

Similarly, we can estimate the TC-loss in the same way:
\begin{equation}
\begin{aligned}
  \Loss_\mathrm{TC} &=
  \int q(\vz) \ln \left(\frac{q(\vz)}{\prod_j q(z_j)}\right) \mathrm{d}z \\
  &= \overline{\ln q(\vz_n)-\ln(\prod_j q(z_{n, j}))}.
  \label{loss:TC}
\end{aligned}
\end{equation}

In practice, as we would like a set of disentangled parameters,
$\Loss_\mathrm{TC}$ would be close to $0$.
In this case, as $q(\vz_n)$ and $\prod_j q(z_{n, j})$ are not perfectly
normalized due to sampling, the difference in their normalization
may result in a negative $\Loss_\mathrm{TC}$.
However, adding the normalization factor into the estimator would
overestimate the TC loss.
As a result, we decided to keep the unbiased form of \autoref{loss:TC}
while keeping in mind that a small negative $\Loss_\mathrm{TC}$
might occur.

\subsection{Cyclical Annealing}
\label{sec:cyc_anl}

A problem with both $\beta$-VAE and $\beta$-TCVAE is that, at the
beginning of training process, the latent $\vz$ can hardly represent the
dataset due to the random initialization.
However, all those KL terms in the ELBO would push the posterior
$q(\vz|\vx)$ to the uninformative prior $p(\vz)$.
As a result, the decoder would tend to do reconstruction while ignoring
the latents, which is usually called ``Latent Space Collapse''.

To deal with this problem, we choose to do cyclical annealing during
training.\cite{fu2019cyclicalannealing}
To be specific, we add a new parameter $\lambda$ to control all the KL
terms in the ELBO and periodically varying it in the whole training
process:
\begin{equation}
  \begin{aligned}
    &\Loss'_{\beta-\mathrm{TC}}\\
    =&\mathbb{E}_{q(\vz|\vx_n)p(\vx_n)}[\log p(\vx_n|\vz)]\\
    &- \lambda[\alpha \KL(q(\vz,\vx_n)||q(\vz)p(\vx_n)) + \beta\KL(q(\vz)||\prod_j q(z_j))
    + \gamma \sum_{j} \KL(q(z_j)||p(z_j))].
  \end{aligned}
  \label{eq:lambda}
\end{equation}

When $\lambda$ is small, the ELBO would be dominated by the
reconstruction loss, so the model will focus on learning to accurately
reconstruct the data, ensuring that the latent variables capture
significant information about the inputs.
On the other hand, during the high $\lambda$ phase, KL terms becomes
important, which helps in learning a disentangled and structured latent
space.
By alternating between these phases, cyclical annealing allows the model
to find a better balance, improving both reconstruction quality and
latent space disentanglement without falling into the extremes of latent
space collapse or under-regularization.

\section{Model Architecture}
\label{sec:en/de-coder}
We utilize a combination of multilayer perceptron (MLP) architecture and
pointwise convolution layers to construct our Variational Autoencoder
(VAE) model, ensuring a symmetric and efficient structure for encoding
and decoding.

The encoder is designed to process $T^2(k, z)$
from three snapshots along with the five
cosmological parameters.
For the first step, a pointwise convolution is applied to each k-bin
independently.
This operation increases the feature dimensionality
at each bin without changing the number of $k$-bins,
allowing the model to extract richer local representations.
The resulting intermediate features are then concatenated
with the five cosmological parameters
and passed into a multi-layer perceptron (MLP).
The MLP then output the mean and variance of the latent variables, forming the posterior distribution.

The decoder takes the sampled latent variables together with the same five cosmological parameters as input.
These inputs are first processed by another MLP, which produces an intermediate representation.
This intermediate output is reshaped to match the number of $k$-bins in the power spectrum
and then passed through a pointwise convolution layer to reconstruct $T^2$.
To incorporate the variational decoder as mentioned in Appendix~\ref{sec:recon_loss},
we also output the variance of the reconstructed
$T^2$, $\sigma_{\theta, j} = \mathrm{MSE}(\vx'_j, \vx_j)$.
This approach allows the model to learn a more flexible and accurate
representation of the data by considering the variance in the
reconstruction process \citep{Rybkin2021simple}.

\begin{table}[tb]
\centering
\caption{All Hyperparameters Tuned by Optuna.
The range for parameters is just for reference.
For a specific training run, the range would be changed, and some
parameters might be fixed for a model with the desired behavior.
For \texttt{ord}, ``B'', ``A'', and ``D'' stand for ``Batchnorm'', ``Activation
function'', and ``Dropout''.
\texttt{lgta} is the leaky slope transformed with the Sigmoid function.
For \texttt{mu} and \texttt{nu}, the number of epochs in each annealing
cycle is $1/(\texttt{mu}\cdot\texttt{nu})$ and the duration of annealing
is $1/\texttt{mu}$.}
\label{table:hpara}
\begin{tabular}{lllll}
\toprule
\textbf{Category} & \textbf{Name} & \textbf{Type} & \textbf{Range} & \textbf{Description} \\
\midrule
\multirow{6}{*}{Model}
  & \texttt{Dc}    & int            & [1, 10]                    & Depth of convolutional layers \\
  & \texttt{2Wc}   & int            & [2, 10]                    & Convolution layer width is $2^{\texttt{2Wc}}$ \\
  & \texttt{Dm}    & int            & [1, 10]                    & Depth of MLP \\
  & \texttt{2Wm}   & int            & [4, 10]                    & MLP layer width is $2^{\texttt{2Wm}}$ \\
  & \texttt{lgta}  & float          & [-5, 5]                    & Logit of leaky ReLU slope \\
  & \texttt{ord}   & categorical    & \{BA, A, AB,  & Module order in architecture \\
  &                &                & BAD, AD, ABD\}& \\
\midrule
\multirow{6}{*}{Loss}
  & \texttt{lambda} & float (log)   & $[10^{-4}, 10^{-1}]$       & Overall KL scaling \\
  & \texttt{mu}     & float (log)   & $[2/\texttt{num\_epoch}, 1]$ & Controls annealing duration \\
  & \texttt{nu}     & float (log)   & $[0.1, 1]$                 & Controls annealing frequency \\
  & \texttt{alpha}  & float (log)   & $[0.1, 10]$                & Weight for MI loss \\
  & \texttt{beta}   & float (log)   & $[0.1, 1]$                 & Weight for TC loss \\
  & \texttt{gamma}  & float (log)   & $[10^{-3}, 10]$            & Weight for dw-KL loss \\
\midrule
\multirow{9}{*}{Training}
  & \texttt{lr}     & float (log)   & $[10^{-6}, 10^{-1}]$       & Learning rate \\
  & \texttt{1-b1}   & float (log)   & $[10^{-3}, 1]$             & $1 - \beta_1$ for AdamW \\
  & \texttt{1-b2}   & float (log)   & $[10^{-6}, 1]$             & $1 - \beta_2$ for AdamW \\
  & \texttt{wd}     & float (log)   & $[10^{-6}, 10^{-1}]$       & Weight decay \\
  & \texttt{inid}   & categorical   & \{uniform, normal\}        & Initialization distribution \\
  & \texttt{inim}   & categorical   & \{fan\_in, fan\_out\}      & Initialization mode \\
  & \texttt{bnmf}   & float (log)   & $[10^{-2}, 10^{2}]$        & BN momentum \\
  & \texttt{dopc}   & float (log)   & $[0.01, 0.5]$              & Dropout rate in CNN \\
  & \texttt{dopm}   & float (log)   & $[0.01, 0.5]$              & Dropout rate in MLP \\
\bottomrule
\end{tabular}
\end{table}

\begin{figure*}[tb]
\centering
\includegraphics[width=0.6\textwidth]{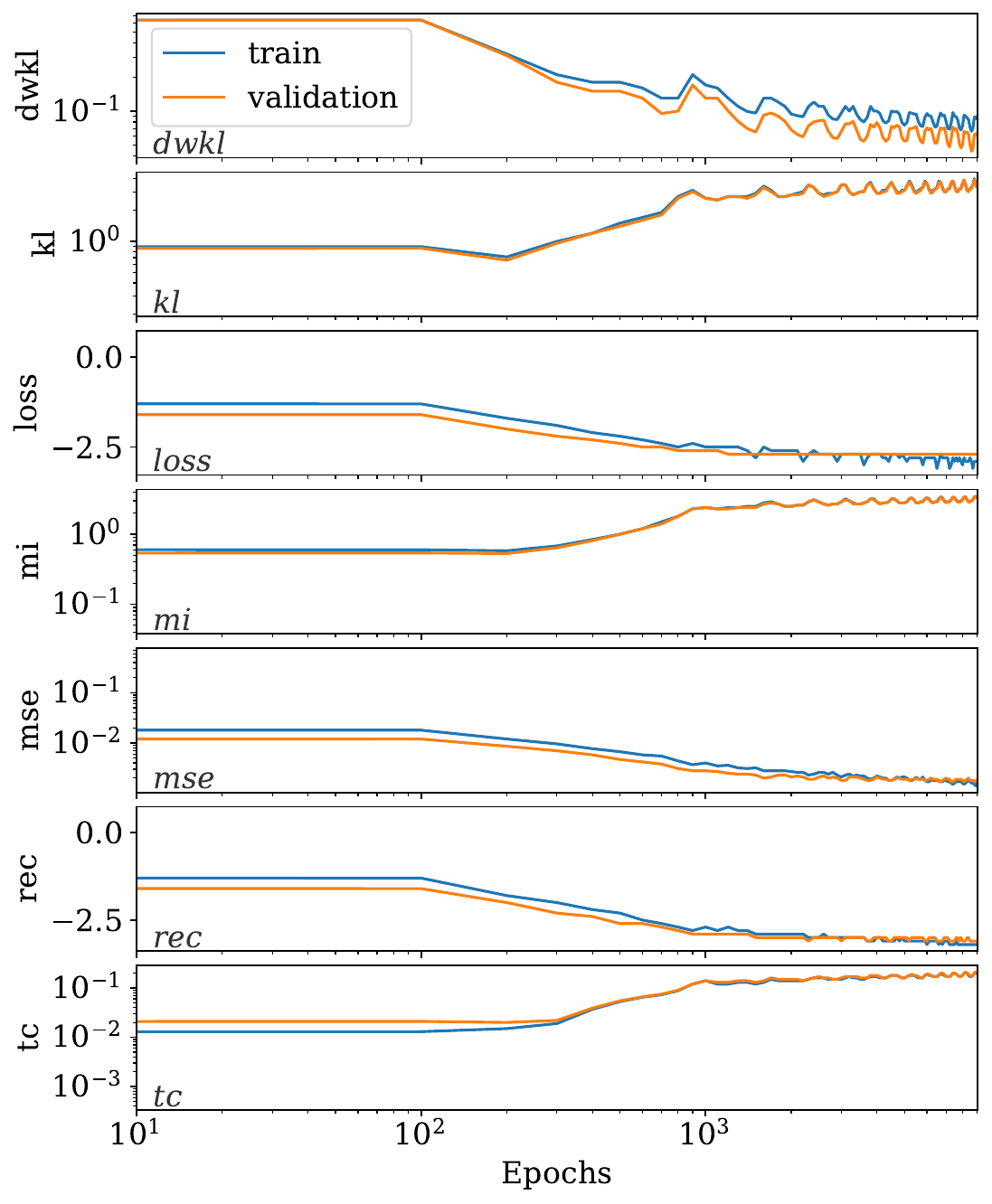}
\caption{An example of the training and validation loss curves.
We use a learning rate scheduler with a patience of 1000 epochs and a
1\% threshold, along with early stopping if no improvement is observed
within 2000 epochs.
Due to the cyclic annealing strategy described in
Appendix~\ref{sec:cyc_anl}, the KL and reconstruction losses may exhibit
oscillations during training, but they generally converge to stable
values by the end.}
\label{fig:loss_curve}
\end{figure*}

\begin{figure}[tb]
  \centering
  \includegraphics[width=0.5\textwidth]{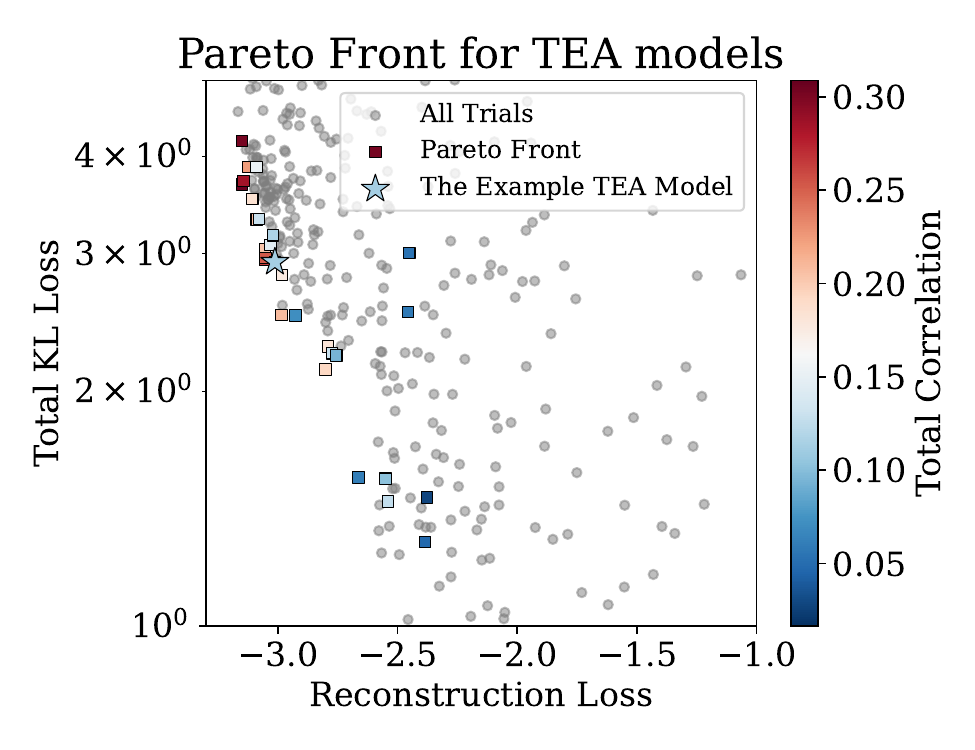}
  \caption{
    Pareto front of TEA models selected during hyperparameter optimization.
    Each point represents a trial in the \texttt{OPTUNA} search space,
    plotted by its reconstruction loss (horizontal axis) and total KL loss (vertical axis),
    with color indicating the total correlation (TC) loss.
    Gray circles denote all trials, colored squares highlight the Pareto-optimal solutions,
    and the blue star marks the final model selected for all the plots shown in the paper.
  }
  \label{fig:pareto_front}
\end{figure}

\section{Model Tuning}
\label{app:hpara}

As noted in \autoref{sec:training},
we optimize 21 hyperparameters related to the model architecture,
training procedure, and loss formulation using the Python package \texttt{OPTUNA}.
A complete list of these hyperparameters is provided in \autoref{table:hpara}.
In addition to the optimized settings,
we fix the batch size to match the size of the training (or validation) set
to avoid bias in loss estimation that can arise from small batches.

\texttt{OPTUNA} searches for optimal hyperparameter configurations
by simultaneously minimizing the reconstruction loss, KL divergence loss,
and total correlation (TC) loss.
To avoid selecting models with degenerate latent posteriors,
we impose the following constraint:

\begin{equation}
1 \leq A = \frac{\mathrm{Var}(\mu(\vx))}{\overline{\sigma(\vx)^2}} \leq 10,
\end{equation}

where $\mu(\vx)$ and $\sigma(\vx)$ denote the mean and standard deviation
of the latent posterior for a given input $\vx$,
and the variance is computed across the entire dataset.
This criterion ensures that the latent representation is
neither too concentrated (i.e., collapsing to a nearly constant value)
nor overly dispersed (i.e., resembling delta functions as in traditional autoencoders).
All models that fail to satisfy this condition are pruned,
and only the remaining models are considered in the later analysis.

Due to the inherent trade-off between reconstruction loss and regularization losses (KL and TC),
the hyperparameter optimization yields a Pareto front rather than a single global minimum.
As illustrated in \autoref{fig:pareto_front}, most Pareto-optimal models lie along
a frontier governed primarily by reconstruction and KL loss,
with a few outliers exhibiting exceptionally low TC loss (depicted as blue squares).
With the Pareto-optimized parameters, we retrain the models using those hyperparameters with a longer training process,
and then select the best models among the retrained ones.
An example of the retraining loss curve is shown in \autoref{fig:loss_curve}.
The final model is selected by balancing reconstruction accuracy with latent disentanglement, and is highlighted by a star in the plot.

\section{Results with SIMBA}
\label{app:EAST}

\begin{figure}[tb]
\centering
\includegraphics[width=0.8\textwidth]{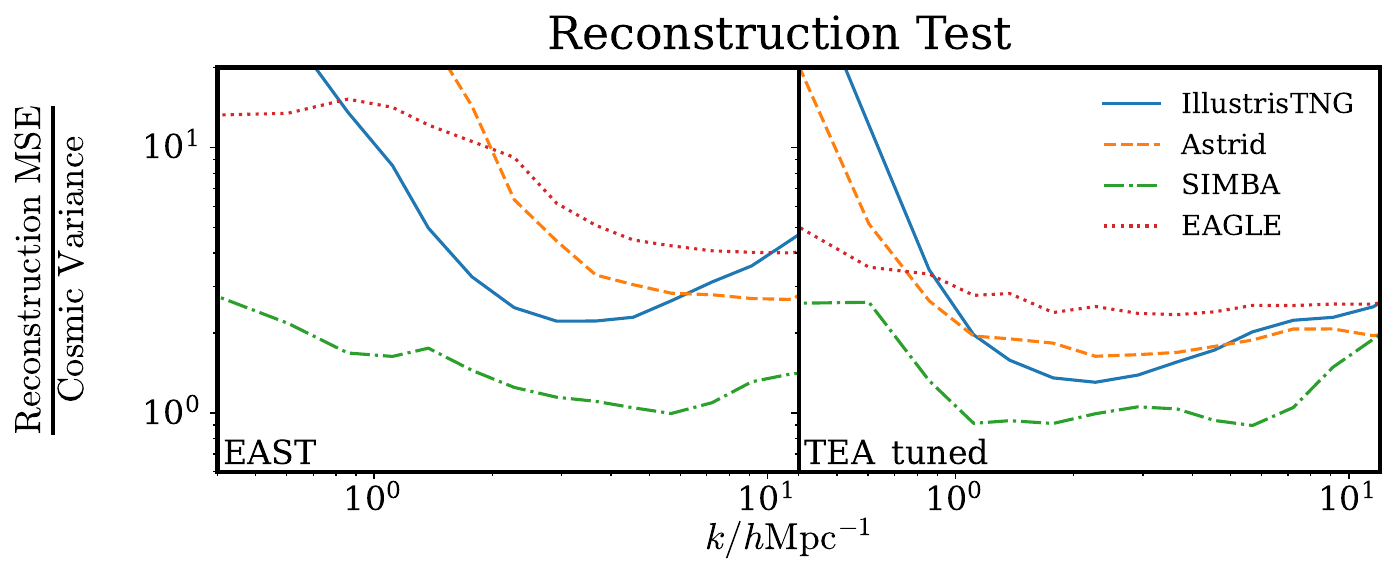}
\caption{
Same as \autoref{fig:chi2_test}, but for the EAST and ``TEA tuned'' models.
While the EAST model performs well on the SIMBA suite,
it exhibits significantly higher variance on the other suites, particularly at low-$k$.
In contrast, the ``TEA tuned'' model achieves consistently better reconstruction across all suites.
}
\label{fig:EAST_chi2}
\end{figure}

\begin{figure*}[tb]
\centering
\includegraphics[width=0.35\textwidth]{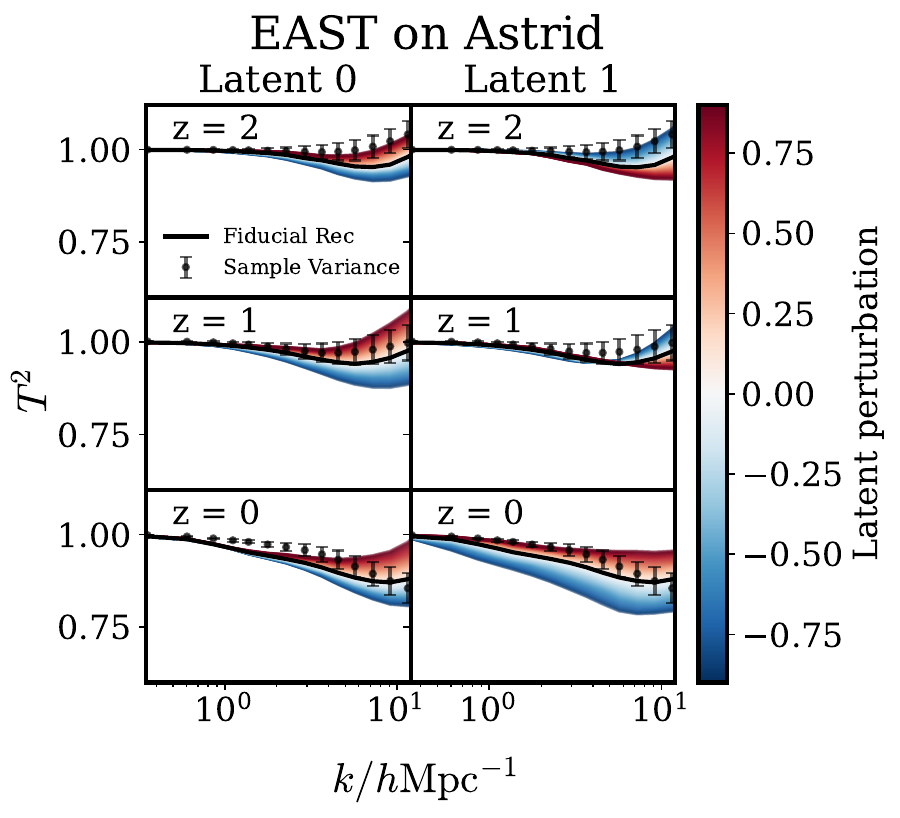}
\includegraphics[width=0.35\textwidth]{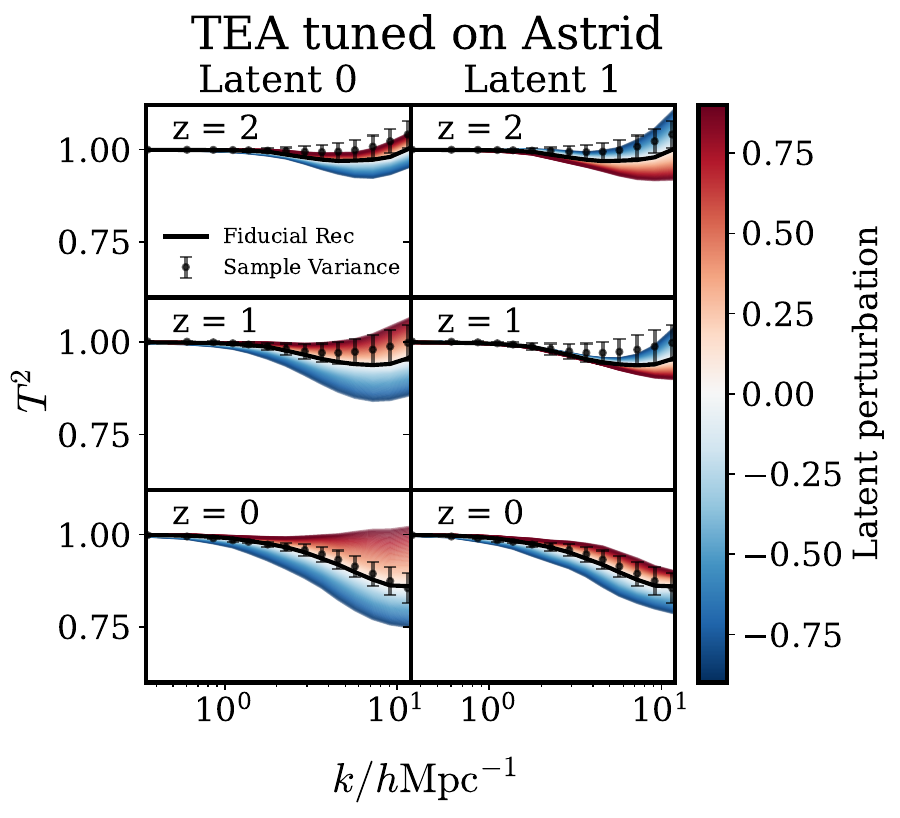}
\caption{
Same as the middle-left panel of \autoref{fig:AIO_plot},
but showing results from the EAST and ``TEA tuned'' models on the Astrid suite.
}
\label{fig:EAST_rec}
\end{figure*}

\begin{figure*}[tb]
\centering
\includegraphics[width=0.5\textwidth]{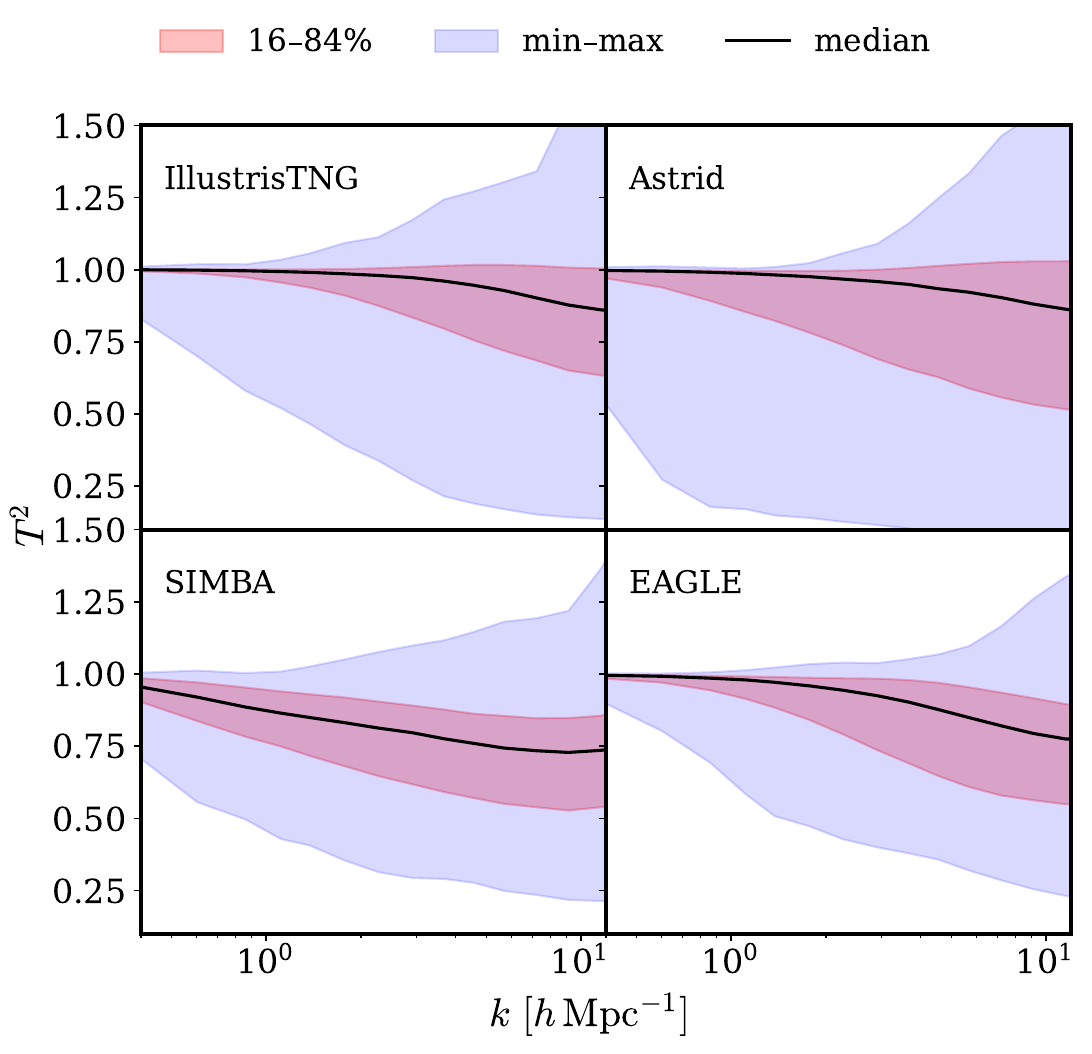}
\caption{
Range of $T^2$ for different suites. The blue region stands for the range between upper and lower limit in each $k$ bin.
The red region represents the $16-84\%$ range in each $k$ bin.
Comparing to the other suites, SIMBA both significantly deviate from 1 and have much larger variance on the largest scales.
}
\label{fig:T2_scatter}
\end{figure*}

As discussed in \autoref{sec:training},
our general training dataset for the TEA model does not include SIMBA LH6 due to its strong baryonic feedback.
In this section, we present results involving the SIMBA suite
to further justify the construction of the ``TEA'' model.

We compare two models. The first, denoted as \textbf{EAST},
is trained on all four simulation suites (IllustrisTNG, Astrid, SIMBA, and EAGLE)
using the same hyperparameter setup described in \autoref{app:hpara}.
The second, called the \textbf{``TEA tuned''} model, starts from the EAST model
and is further fine-tuned on the TEA dataset, which excludes SIMBA.

Figure~\ref{fig:EAST_chi2} shows reconstruction performance on the cross-validation set.
The EAST model achieves excellent reconstruction on SIMBA
but performs poorly on the other three suites, especially at large scales (low-$k$),
where the variance is much larger.
In contrast, the ``TEA tuned'' model exhibits significantly improved performance
on the other suites while maintaining acceptable behavior on SIMBA.

This issue is further illustrated in \autoref{fig:EAST_rec},
which mirrors the latent variation analysis shown
in the middle-left panel of \autoref{fig:AIO_plot}.
The EAST model shows strong reconstruction bias at low $k$ in the Astrid suite,
and the latent dimensions exert a large influence in that regime.
After fine-tuning on the TEA dataset, the ``TEA tuned'' model significantly reduces this bias,
particularly in the large-scale regime.
A reason for this is that, SIMBA not only have a stronger feedback than the other suite on the largest scales, but also have larger variance due to the wide range of CAMELS, as shown in \autoref{fig:T2_scatter}.
Based on these results, we eventually decide to use TEA model for the main analysis shown in this paper.

\section{Reproducibility Test}

\begin{figure}[tb]
\centering
\includegraphics[width=0.7\textwidth]{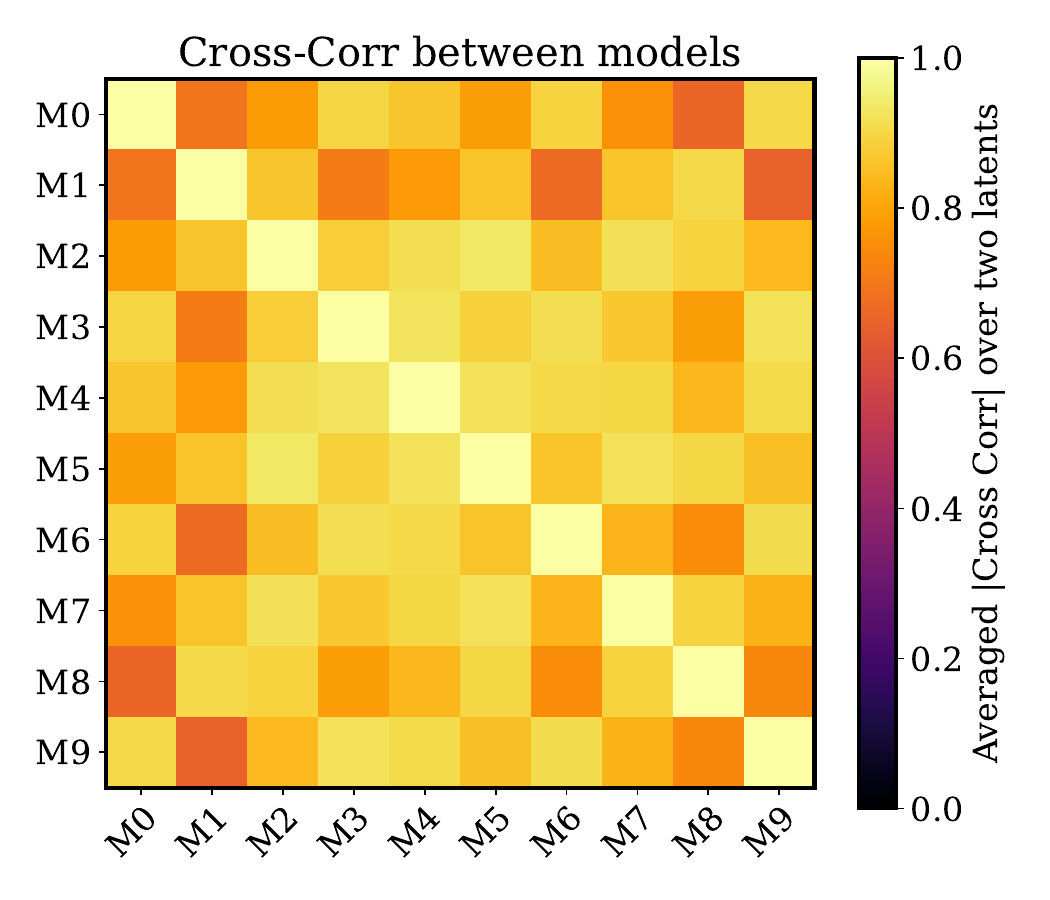}
\caption{Cross-correlation matrix between latent space from 10
independently trained TEA models, where ``M0'' is the one we choose for
the main result of the paper.
Each model is trained with identical hyperparameters on combined data
from four simulation suites (IllustrisTNG, Astrid, SIMBA, and EAGLE).
To account for the inherent symmetry in the latent space, we align each
model pair by evaluating all possible axis permutations and sign flips,
reporting the maximum average absolute Pearson correlation across the
two latent dimensions.
All model pairs exhibit strong correlations ($\geq 0.6$), with a mean of
0.84, indicating robust reproducibility of the learned latent space
despite stochastic variations in training.
}
\label{fig:reproducibility}
\end{figure}

Due to the stochastic nature of the VAE training process,
we assess reproducibility by training 10 models with identical hyperparameters
and analyzing the cross-correlation between their latent representations.
In \autoref{fig:reproducibility},
we present the average absolute cross-correlation across all 10 models,
evaluated on the combined data from the four simulation suites.
The ``M0'' model is the ``TEA'' model mentioned in the main text.
To account for the inherent symmetry in latent space,
we consider all possible axis permutations and sign flips,
and report the maximum correlation for each model pair.
All model pairs exhibit cross-correlations above 0.6, with a mean value of 0.84,
demonstrating a high degree of consistency
across retrainings and the robustness of our latent space.

\section{PCA Reconstruction Test}
\label{app:PCA}

\begin{figure}[tb]
\centering
\includegraphics[width=0.8\textwidth]{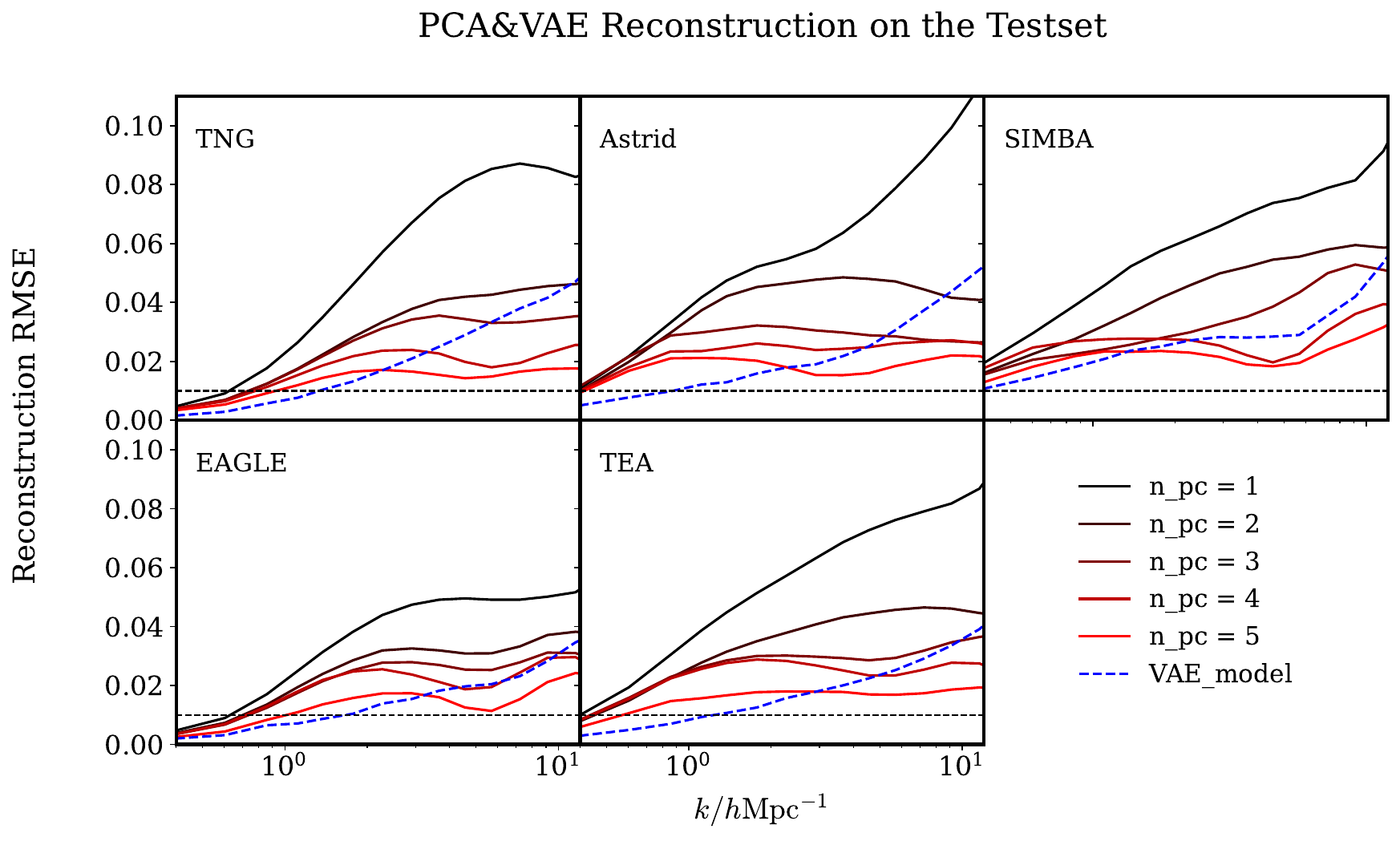}
\caption{
Reconstruction RMSE of matter power spectrum ratios on the test set for our VAE models,
compared with PCA using varying numbers of principal components (PCs).
Each panel corresponds to one simulation suite: IllustrisTNG, Astrid, SIMBA, EAGLE, and a combined ``TEA'' training case.
Here PCA is trained on the same training data as the corresponding VAE.
In the ``TEA'' panel,  PCA is trained on the TEA training set for comparison to the reconstruction of the TEA model, and we show reconstruction performance on the joint test set across the three suites of TEA (IllustrisTNG, EAGLE, and Astrid).\\
Solid curves show PCA reconstruction errors for $n_{\rm pc}=1\sim5$, with darker colors indicating fewer components, while the blue dashed curve shows the VAE reconstruction.
The horizontal dashed line marks the 1\% RMSE threshold.\\
Across all scales, our 2-dimensional VAE achieves better reconstruction than PCA with 2 PCs and performs comparably to PCA with 3 PCs.
Notably, the VAE outperforms PCA with 5 PCs on large scales for all suites.
Moreover, around $k = 1\,h/\mathrm{Mpc}$, PCA reconstruction rarely reaches the 1\% RMSE level even with 5 PCs, while the VAE meets or approaches this accuracy for all cases except for SIMBA.
}

\label{fig:PCA_rec}
\end{figure}

To evaluate the capability of our VAE models versus the conventional
linear principal component analysis (PCA) approach \citep{Eifler2015, Huang2019bar},
we compare their reconstruction root-mean-square errors (RMSE)
across simulation suites.
The RMSE is defined as: $\mathrm{RMSE}(T^2) = \sqrt{\overline{(T'^2-T^2)^2}}$.

For a fair comparison,
PCA is trained using the same training data as the corresponding VAE model
for each suite (IllustrisTNG, Astrid, EAGLE, SIMBA, and TEA).
The principal components are obtained by diagonalizing the covariance matrix of $T^2$.
The resulting reconstruction errors are shown in \autoref{fig:PCA_rec}.

As illustrated in the figure,
our 2-dimensional VAE models consistently outperform PCA with two PCs and achieve comparable performance to PCA with three PCs.
On large scales, the VAE even surpasses PCA with five PCs across all suites.
In particular, at $k \approx 1\,h/\mathrm{Mpc}$, the VAE reconstructs the transfer function with $\sim$1\% RMSE,
whereas PCA fails to reach this accuracy in all suites.

This performance gap is expected because PCA captures only the leading linear modes of variation in the transfer function,
while the VAE learns a nonlinear latent representation.
As a result, PCA requires multiple PCs to reach the same accuracy at the same $k$ and has to sacrifice low $k$'s and likely some parameter choices,
whereas the VAE can efficiently encode and reconstruct these effects using only two latent variables.

\bibliographystyle{aasjournalv7}
\bibliography{repr}

\end{document}